\documentclass[10pt]{article}

\usepackage[a4paper, margin=0.8in]{geometry}

\providecommand{\keywords}[1]{\textbf{\textit{Keywords---}} #1}

\usepackage{epstopdf}
\usepackage{subfigure}

\usepackage{bbm}
\usepackage{dsfont}

\usepackage{changepage}
\usepackage[normalem]{ulem} 

\usepackage{algorithm}
\usepackage{algpseudocode}

\usepackage{amsmath}
\usepackage{booktabs}
\usepackage{multirow}
\usepackage{changepage}

\usepackage{authblk}

\usepackage{url}

\usepackage{todonotes}

\begin{document}


\title{\bf A high-frequency approach to Realized Risk Measures}

\author[1]{Federico Gatta}
\author[1,2]{Fabrizio Lillo}
\author[3]{Piero Mazzarisi\thanks{Corresponding author. Email: piero.mazzarisi@unisi.it}}

\affil[1]{Scuola Normale Superiore, Pisa, Italy}
\affil[2]{University of Bologna, Italy}
\affil[3]{University of Siena, Italy}


\date{}

\maketitle

\begin{abstract}
We propose a new approach, termed  {\it Realized Risk Measures} (RRM), to estimate Value-at-Risk (VaR) and Expected Shortfall (ES) using high-frequency financial data. It extends the {\it Realized Quantile} (RQ) approach proposed by Dimitriadis and Halbleib \cite{dimitriadis2022realized} by lifting the assumption of return self-similarity, which displays some limitations in describing empirical data. More specifically, as the RQ, the RRM method transforms intra-day returns in intrinsic time using a subordinator process, in order to capture the inhomogeneity of trading activity and/or volatility clustering. Then, microstructural effects resulting in non-zero autocorrelation are filtered out using a suitable moving average process. Finally, a fat-tailed distribution is fitted on the cleaned intra-day returns. The return distribution at low frequency (daily) is then extrapolated via either a characteristic function approach or Monte Carlo simulations. VaR and ES are estimated as the quantile and the tail mean of the distribution, respectively. The proposed approach is benchmarked against the RQ through several experiments. Extensive numerical simulations and an empirical study on 18 US stocks show the outperformance of our method, both in terms of the in-sample estimated risk measures and in the out-of-sample risk forecasting.
\end{abstract}



\section{Introduction}
\label{sec:intro}
Risk measures play an essential role in the financial industry and market regulation. Value at Risk (VaR) and Expected Shortfall (ES) are the most widely used risk measures. The former is defined as the quantile of the return distribution and the latter is the tail mean, i.e., the mean computed over the tail below the quantile. More formally, let $\theta\in[0,1]$ be a probability level and assume that we are at time $t=0$ with a horizon $T>0$. 
Let $(\Omega, \mathcal{F}, \mathds{P})$ be a probability space and
$Y$ be the random variable describing the asset log returns, whose Cumulative Density Function (CDF) is $F_Y$, with quantile function $F^{-1}_Y$. Then, VaR and ES at level $\theta$ are defined as:
\begin{equation}
    VaR_\theta(Y) = F_Y^{-1}(\theta) \quad\quad and \quad\quad ES_\theta(Y) = \mathds{E}[Y | Y \le VaR_\theta]
\end{equation}

Due to its simplicity, VaR has been the most widely used risk measure in recent years. However, due to changes in regulatory policy (see \cite{feridun2020basel} for a reference) and theoretical advancements \cite{fissler2016higher}, ES has been suggested as a new standard in risk management to be preferred to VaR.
In fact, it offers a more comprehensive view of the risk as it contains information on the whole tail of the asset return distribution. Moreover, it has better mathematical properties, in particular it is sub-additive, thus ensuring that the corresponding portfolio optimization problem is well-posed. This has recently driven academic effort towards the estimation and forecasting of such risk measures, especially at  daily frequency.
Noticeably, several papers addressing VaR and/or ES forecasts have been proposed in the general framework of AutoRegressive (AR) processes \cite{gatta2024caesar, patton2019dynamic, taylor2019forecasting} and deep learning models \cite{barrera2022learning}. However, the problem presents many challenging aspects due to the intrinsic difficulty in estimating the tail properties of a distribution due to data scarcity. Moreover, the backtesting itself remains a challenging task, 
as pointed out in \cite{acerbi2014back}, thus jeopardizing the comparison among different predictive methods.\\ 

Leveraging high-frequency financial data (intra-day returns) for a more precise estimation of VaR and ES at low-frequency (daily time scale) is the topic addressed in this paper.
Realized Volatility (RV) is a clear reference for devising a methodology using high-frequency data to estimate an aggregated quantity at low-frequency. Inspired by this, we
refer to our proposed methodology as {\it Realized} Risk Measures (RRM). However, notice that the parallelism between RRM and RV is limited to the passage from high-frequency data to low-frequency estimates. Indeed, realized volatility approximates a latent variable described by a stochastic process and the only assumption is the existence of the second-order moment of the process. Instead, VaR and ES are tail properties of the return distribution, and, as such, a distributional assumption stronger than the RV approach is required for estimation.
Finally, it is worth noting that, although the underlying idea is quite simple, to the best of our knowledge, there is only one paper taking this direction, namely the seminal contribution by Dimitriadis and Halbleib \cite{dimitriadis2022realized}. However, as discussed in the following, it is strongly based on the assumption of self-similarity of the log-price process measured in intrinsic time. As we show below, empirical high frequency return time series deviates from self-similarity, thus challenging the applicability of the model of Ref. \cite{dimitriadis2022realized}.\\

The general pipeline of our proposal can be summarized as follows. We estimate the realized risk measures in each day separately. In the following, we consider time $t=0$ as the opening of the regular trading hour of a given day, that is 09:30 at NYSE and NASDAQ considered in the empirical analysis below, and final time $T$ as the end of the day, that is the closing time 16:00 of the market. We focus on the minute-by-minute data on the regular trading hours, thus discarding any information from the overnight period. We collect 390 minute observations for each day. Let $\{S^{(t)}_i\}_{i=0}^{390}$ be the intra-day log-price series for a specific day $t$. In the following, we omit the subscript $(t)$ when there is no risk of confusion. Let $Y:=S_{390}-S_0$ be the daily log return defined as the difference between the closing and opening prices. The goal is to estimate VaR and ES associated with the random variable $Y$.
The proposed methodology provides a solution in a number of steps. First, a coarse-graining procedure is considered based on a subordinator allowing to define unequally-spaced time increments over which we aggregate data, in order to capture the variability in the intensity of market activity, see \cite{mandelbrot1997multifractal,dimitriadis2022realized}.
The intuition behind this non-homogeneous aggregation is cleaning from possible intra-day effects and making the return process more similar to a Brownian motion.
In the second step, a pre-processing procedure based on a Moving Average (MA) filter is applied to remove microstructural effects, resulting in a non-zero autocorrelation at lag one. Then, under the assumption that MA residuals are iid, a fat-tailed distribution is fitted. Finally, the last step involves a scaling procedure from an intra-day to a daily timescale. This can be achieved in two different ways: (i) scaling to the daily frequency via a characteristic function approach in the Fourier space, obtaining numerically the desired distribution function based on the Gil-Pelaez theorem; (ii) scaling to the daily frequency via Monte Carlo simulations. Since the two methods have pros and cons, we finally consider their equally weighted ensemble average to reduce the estimation error.\\

From the above description, it is clear that the most critical assumption is that the subordinated and cleaned returns are iid. This might lead to misestimation of VaR and ES. As a reference point, we compare our results with those obtained using the approach of Realized Quantile of \cite{dimitriadis2022realized}. In this case, the main assumption is that subordinated log-price process is self-similar with stationary increments\footnote{Specifically, the process $\{S_{\tau(j)}\}_{j=0}^c$ is said to be self-similar if $\exists H \in (0,1)$ (also known as the Hurst exponent) such that $S_{\tau(j)\Delta} = \Delta^H S_{\tau(j)}$. When the increments of a self-similar process are stationary, then they scale as $\Delta^H$, as well as the VaR and ES. Thus, $VaR(Y)$ can easily be computed as $c^H\; VaR(Y_1)$.}. As we show below, the self-similarity assumption of subordinated prices poorly works in the presence of drift and short-range correlations, as observed in empirical data.
Thus, the comparison between our proposed methodology and the Realized Quantile approach can be seen as a comparison between two approximations: (i) subordinated and cleaned returns are iid (our assumption) versus (ii) subordinated log-prices are self-similar (\cite{dimitriadis2022realized}). An extensive battery of numerical simulations and empirical analyses shows that the former outperforms the latter.\\

The rest of this paper is organized as follows. Section \ref{sec:lit} contains an overview of the literature on risk measures, with a particular focus on the few works proposing how to use high-frequency data for estimation and the associated limitations, providing the motivation for the presented research.
Our proposal is discussed in Section \ref{sec:met}, together with the backtesting criteria adopted for evaluation. Section \ref{sec:exp} presents the experiments corroborating our methodology. Section \ref{sec:conc} concludes and discusses possible extensions. Some technical Appendices present implementation details, additional simulation, and empirical results. Appendix \ref{app_imp_det} discusses technical implementation details. Appendix \ref{app:loss_bias} contains additional insights into the simulation results. Finally, Appendix \ref{app:for_rrm} presents more out-of-sample experimental results.


%
%
\section{Risk Measurement and  Limitations of Existing Methods}
\label{sec:lit}
This section examines the two primary approaches for measuring and forecasting financial risk. Additionally, we highlight how specific types of data, i.e., on daily or high-frequency time scales, can influence the choice of the modeling approach.

%
\subsection{Risk Forecasting Based on Daily Data}
Risk forecasting is a central challenge in quantitative finance, as it is critical in risk management, portfolio optimization, and derivative pricing. Among risk measures, volatility has received much attention throughout the years, thus representing a standard reference framework for any risk forecasting approach; see \cite{andersen2006volatility} for a review. In the univariate case, with observations available at discrete points in time, let us refer to $y^{(t)},\:\:t=1,2,\ldots$ as the daily sampled observations of the logarithmic returns process\footnote{We indicate the day index in superscript brackets, and the intra-day one in the subscript.} $\{Y^{(t)}\}_{t=1,2,\ldots}$. Assuming the existence of the conditional second moments of the process, let us define $\mu_{t+1\vert t}=\mathds{E}[Y^{(t+1)}\vert \mathcal{F}_t]$ and $\sigma_{t+1\vert t}^2=\mbox{Var}[Y^{(t+1)}\vert \mathcal{F}_t]$ as the mean and variance of $Y^{(t+1)}$ conditional to filtration $\mathcal{F}_t$ reflecting all relevant
information up to time $t$. The volatility $\sigma_{t+1\vert t}$ can be viewed as a basic risk measure, and it is typically modeled with a model of the GARCH family (introduced in \cite{engle1982autoregressive} and \cite{bollerslev1986generalized}) and the discrete-time process is estimated via maximum likelihood.\\

However, the Basel framework has favored the diffusion of a new standard for capital requirement based on Value-at-Risk and, more recently, on Expected Shortfall as measures of daily risk. To this end, \cite{koenker1978regression} provided a general regression framework for risk forecasting by proving that the quantile (or, equivalently, the VaR) $\hat{q}$ at probability level $\theta$ can be consistently estimated by minimizing the Pinball loss $\mathcal{L}^\theta_q(\hat{q}, Y)$ defined as
\begin{equation}
    \label{eq:1113_0900}
    \mathcal{L}^\theta_q(\hat{q}, Y) := (Y - \hat{q})\left( \theta - \mathbbm{1}_{\{Y\le\hat{q}\}} \right)
\end{equation}
where $\mathbbm{1}$ is the indicator function. Inspired by the GARCH volatility modeling, Ref. \cite{engle2004caviar}  proposed the CAViaR approach to model heteroskedastic effects for $\hat{q}$. The VaR at future time steps is regressed against a non-linear transformation of the observed returns and the previous quantile estimates. The process is then estimated by minimizing the Pinball loss.\\

The main drawback of VaR is that it is not a coherent risk measure since it is not sub-additive (see, e.g., \cite{artzner1999coherent}), thereby discouraging diversification and posing serious limitations to its usage for portfolio problems. ES is instead a coherent risk measure and it became more and more used in risk management, see \cite{acerbi2002expected,acerbi2002coherence}. Interestingly, \cite{fissler2016higher} proved that VaR and ES can be jointly estimated as the minimum of a specific class of loss functions. One of the most used in the literature is  
\begin{equation}
    \label{eq:1113_0901}
    \mathcal{L}^\theta_e(\hat{q}, \hat{e}, Y) := \frac{\hat{q}}{\hat{e}} - \frac{\hat{q} - Y}{\theta\hat{e}} \mathbbm{1}_{\{Y\le\hat{q}\}} + log(-\hat{e})
\end{equation}
where $\hat{e}$ is the ES estimate. This loss allows the introduction of a regression framework similar to GARCH and CAViaR for ES forecasting.
In particular, the CAESar model by \cite{gatta2024caesar} generalizes the CAViaR framework to jointly predict VaR and ES. A different autoregressive approach based on the Generalized Autoregressive Score (GAS) model is in \cite{patton2019dynamic}. Another line of studies is focused on deep learning models. Regarding quantile forecast, it is worth highlighting the Quantile Regression Neural Network (QRNN) paradigm, first introduced in \cite{taylor2000quantile}. The same tool is used in \cite{keilbar2022modelling} for multiasset analysis to extract a network for systemic risk. Finally, \cite{barrera2022learning} proposes a two-step neural estimator for the pair (VaR, ES).
%
\subsection{Risk measures based on high-frequency data}

The availability of intraday high-frequency data makes it possible to define a richer information set to improve risk estimation and forecasting. Realized volatility as a noisy proxy of volatility is the best known example. When the return $Y$ is obtained as a function of an underlying continuous time stochastic process for the log-price $S_t$ with general drift $\mu(t)$ and diffusion $\sigma(t)$ coefficients, it is possible to show that squared volatility is approximated by the expectation of the integrated variance of the stochastic process, namely $\sigma_{t+1\vert t}^2\simeq \mathds{E}\left[ \int_t^{t+1}\sigma^2(s)ds\big\vert \mathcal{F}_t\right]=\mathds{E}\left[\mbox{IV}_{t+1}\big\vert \mathcal{F}_t\right]$, and, in turn, the integrated variance $\mbox{IV}_{t}$ can be consistently approximated by the realized variance, which is defined as the sum of (ex-post) squared intraday returns $\mbox{RV}_{t}=\sum_{j=1}^{390/\Delta}\left(S^{(t)}_{j\Delta}-S^{(t)}_{(j-1)\Delta}\right)^2$, with $\Delta$ the time resolution; see, e.g., \cite{andersen2006volatility} for a review. Modeling the realized volatility directly, as in the well-known Heterogenous AutoRegressive (HAR) model by \cite{corsi2009simple}, has proved to broadly outperform any other approach describing a latent process for $\sigma_{t+1\vert t}$ as in the GARCH model.\\

Building on this, several works have proposed enhancing regression models for VaR using realized volatility as an extra regressor, thus leveraging high-frequency data for better forecasting of the risk measure. 
Ref. \cite{clements2008quantile} proposes a general framework for quantile forecasting based on the idea of pairing a realized volatility forecast, typically obtained via the HAR model, with a mapping to the desired quantile. The specific mapping is defined according to a distributional assumption. For example, \cite{clements2008quantile} assumes a normal distribution while \cite{watanabe2012quantile} use a t-distribution, displaying better results in risk forecasting for the S\&P 500 index as explained by the fat-tails of financial returns.
Ref. \cite{vzikevs2015semi} proposed a linear regression model for quantile forecasting, whose regressors are realized measures for the volatility and the jump components. Interestingly, the empirical results raise serious doubts about the importance of the jump component within this context. Indeed, the fitted coefficients associated with jumps are almost always below the 5\% significance threshold. An extension has been proposed in \cite{kawakami2023quantile}, whose focus is forecasting the conditional quantiles of Bitcoin using multiple realized volatility measures associated with different assets as regressors. Finally, \cite{bee2018realized} generalized the CAViaR model \cite{engle2004caviar} by introducing realized volatility as an additional regressor in quantile forecasting.\\


Two main limitations characterize the general approach proposed in the aforementioned works. First, a distributional assumption is required to describe the tail properties of financial returns\footnote{Notice that this is stronger than assuming the existence of the conditional second moment of a process like in the RV case.}. Second, given a distribution, the dynamic characteristics of the tail are controlled entirely by the second moment. While an assumption on the distribution of returns is needed, dynamic patterns for the tail, such as time-varying kurtosis, should be considered in general for a more precise estimation. In order to overcome the last issue, high-frequency data can be used for a direct estimation of VaR and ES, similar to how realized volatility is a proxy for the integrated (over a day) variance. Then, a model like an autoregression for the {\it realized} risk measure can be used for forecasting, similarly to the HAR model for the realized volatility.\\

To the best of our knowledge, the only work that directly uses high-frequency information to estimate the low-frequency quantile has recently been proposed by \cite{dimitriadis2022realized}. Specifically, the minute-by-minute data are used to estimate the daily VaR and ES. Their framework is similar to our proposal below, which is also sketched in Section \ref{sec:intro}. The starting point is the time series of 390 minute-by-minute observations. They are aggregated by using a subordinator that considers the market activity. A subordinator is a transformation of the clock time, that is, an injective map $\tau:[0,c] \rightarrow [0,390]$, with $c\in\mathds{N}$ (in the empirical analysis, we consider values between 39 and 130), $\tau(0)=0$, and $\tau(c)=390$. Then, $c+1$ observations are obtained, namely $\{S_{\tau(j)}\}_{j=0}^c$, and the corresponding log returns can be computed, i.e. $\{Y_j\}_{j=1}^c$, with $Y_j = S_{\tau(j)} - S_{\tau(j-1)}$. The daily return is, by construction, $Y=\sum_{j=1}^c Y_j$. Four subordinators are tested, and the best performance is almost always obtained with Tri-Power Variation (TPV). More details are provided in Section \ref{sec:met}. Then, the subordinated log-price process $\{S_{\tau(j)}\}_{j=0}^c$ is assumed to be self-similar, with a Hurst exponent $H\in(0,1)$ and stationary increments. This implies that $S_{\tau(j)+\Delta} - S_{\tau(j)} \sim \Delta^H \left[ S_{\tau(j)+1} - S_{\tau(j)} \right]$. As a consequence, the quantile (and ES) is scaled to the daily timescale as $VaR_\theta\left( Y\right) = c^H VaR_\theta\left( Y_1 \right)$. Thus, the intra-day risk measures are empirically estimated, and in order to scale them to the daily frequency, it is sufficient to multiply by $c^H$. Finally, authors empirically find that setting $H=1/2$ outperforms scalings obtained with an empirically estimated Hurst exponent in financial data.
\subsection{Testing for self-similarity and its consequences}
Our work is mainly inspired by the analysis carried out in \cite{dimitriadis2022realized}, mentioned above, which is innovative, elegant, and, at the same time, computationally simple. However, the assumption of self-similarity with $H=1/2$ (as used in \cite{dimitriadis2022realized}) is quite strong and restrictive, as it implies a series of properties that are not empirically verified. In particular, in a self-similar process with $H=1/2$  and stationary increments, it must hold that: (i) the process is monofractal, (ii) the increments are linearly uncorrelated, and (iii) the mean of the process is zero at all times. As we demonstrate below, these conditions are not empirically supported.

First, self-similarity with stationary increments implies that the log-price process is monofractal\footnote{This directly follows from the fact that $S_{\tau(j)+\Delta} - S_{\tau(j)}$ is distributed as $\Delta^H \; \left( S_{\tau(j)+1} - S_{\tau(j)} \right)$, which is independent on $\tau(j)$.}. Mathematically, a process is said to be multifractal\footnote{The definition of monofractality and multifractality is typically given in physical time. Here we present it in subordinated time, since this is what is done in \cite{dimitriadis2022realized}.} if 
\begin{equation}
    \label{eq:1117_1000}
    m(q, \Delta) := \mathds{E}\left[ | S_{\tau(j)+\Delta} - S_{\tau(j)} |^q \right] = A(q) \Delta^{H(q)} \quad q, \Delta > 0
\end{equation}
The function $m(q, \Delta)$ is known as the structure-function. Specifically, the above equation implies
\begin{equation}
    \label{eq:1117_1001}
    \log\left( \mathds{E}\left[ | S_{\tau(j)+\Delta} - S_{\tau(j)} |^q \right] \right) = \log\left( A(q) \right) + H(q) \log \Delta
\end{equation}
When the function $H(q)$ is linear, that is, $H(q)=Hq$, the process is monofractal with Hurst exponent $H$. A common testing procedure is to fix a value of $q$, estimate the expected values in Eq. \eqref{eq:1117_1001} using sample averages, for different values of $\Delta$, take the logarithm and regress it against $\log \Delta$, to finally get the value of $H(q)$ (for further references, see \cite{bouchaud2000apparent} and \cite{livieri2018rough}). By repeating the experiment for many different values of $q$, one can study the curve $H(q)$ and test the linearity of the relation. 

\begin{figure}[t]
	\centering
	\includegraphics[scale=0.45]{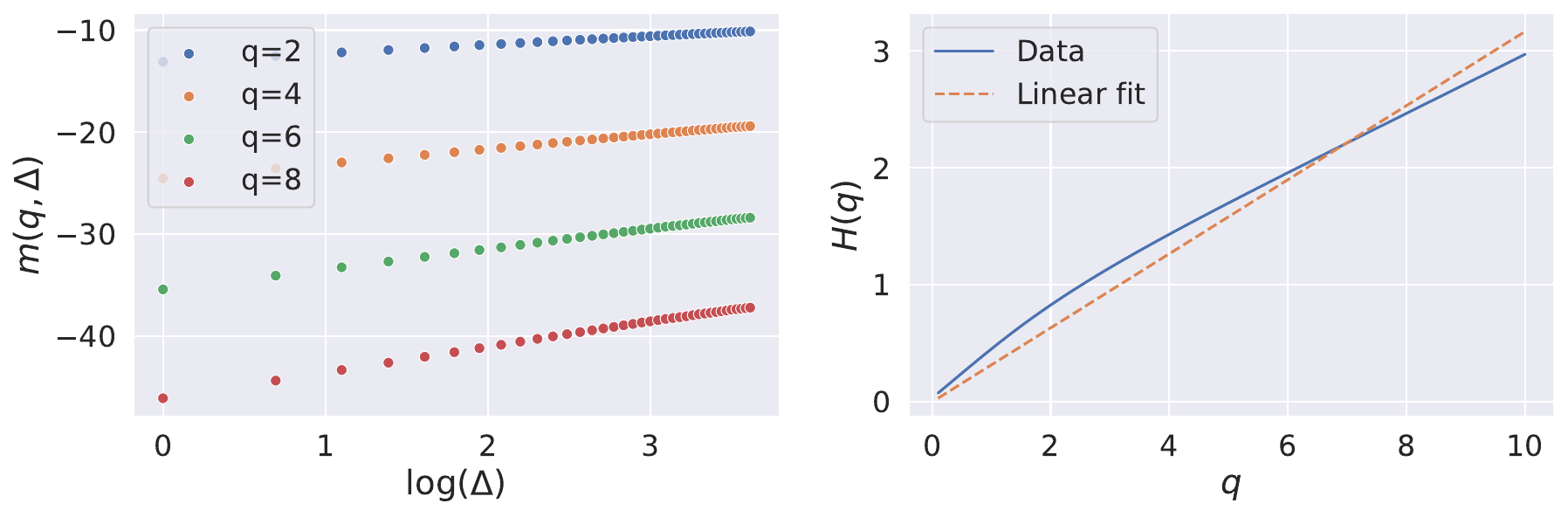}
	\caption{Bank of America (BAC) stock. TPV-based subordinator with $c=130$. The left side of the plot shows the fit $\log\left( \mathds{E} \right)$ against $\log(\Delta)$ for different values of $q$. There is a linear relationship; thus, the log-price process $\{S_{\tau(j)}\}_{j=0}^c$ is at least multifractal. The right side of the plot shows the curve $H(q)$. It is not linear, so we can conclude the process is not monofractal. The plots are obtained by averaging the daily results from 1998 to 2020.}
	\label{img:sf_base}
\end{figure}

We tested monofractality on the dataset described in Section \ref{sec:dataset} and here we present the results for one representative example. We compare the subordinators described in Section \ref{sec:met} and three values of $c$: $39, 78, 130$, which in clock time correspond to 3, 5, and 10-minute frequencies. Following the papers mentioned above, we consider $q\in(0,10)$ and $\Delta\in[1, 38]$ and we use the TPV-based subordinator with $c=130$. Figure \ref{img:sf_base} presents the structure function and $H(q)$ for the stock Bank of America (BAC). These quantities are computed on a daily basis from 1998 to 2020 and then averaged. From the right panel, it is clear that $H(q)$ is a nonlinear function of $q$, indicating that the subordinated price process is not monofractal.\\

A second consequence of the assumption of \cite{dimitriadis2022realized} of a self-similar process with stationary increments with $H=1/2$ is that the process must have uncorrelated increments. 
On the contrary, empirical data display a non-zero autocorrelation, likely due to microstructural effects. To test for it, we have applied the Ljung-Box test \cite{ljung1978measure} (with 5\% threshold) to the subordinated returns time series $\{Y_j\}_{j=1}^c$ obtained with different subordinators and $c$ values. Clearly, the rejection rate that we empirically find depends on the used subordinator, the considered asset, and the selected $c$ value. However, we find that it is above 10\% for every subordinator, suggesting the persistence of an autoregressive component (at the first lag), even after applying the subordinator. The value $H=1/2$ is the one that Ref. \cite{dimitriadis2022realized} finds best fits the data. One could choose $H>1/2$, but this leads to long-range dependence, which is not suitable for the considered type of data. Moreover, the condition $H<1/2$ is very unstable in the presence of noise \cite{beran2017statistics}. The empirical observation of  \cite{dimitriadis2022realized} that the best choice is $H=1/2$ is likely because the bias introduced by neglecting the small autocorrelation at first lags is less detrimental than assuming long-range dependencies. \\

Finally, self-similarity and stationary increments imply zero mean for the stationary increments, at all times (see Theorem 1.3(iii) in \cite{embrechts2000introduction}). If the data have non vanishing mean, the application of scaling arguments leads to inconsistent estimate of the quantile. A simple way to see this is to consider intra-day log-returns, which are iid and distributed as a Gaussian with mean $\frac{\mu}{c} \neq 0$ and standard deviation $\frac{\sigma}{\sqrt{c}}$. Then, the high-frequency quantile reads $q_{HF} = \frac{\mu}{c} + \frac{\sigma}{\sqrt{c}}\alpha$, where $\alpha$ is the standard normal quantile. The true daily quantile is $q_D = \mu + \sigma\alpha$. Assuming no finite-sample estimation errors, the empirical high-frequency quantile equals the true one: $\hat{q}_{HF} = q_{HF}$. Nonetheless, when scaling to the daily timescale one finds $\hat{q}_D = \frac{\mu}{\sqrt{c}} + \sigma\alpha \neq q_D$. A similar reasoning holds for the ES estimator. The empirical results in Section \ref{sec:exp} corroborate the presence of a large bias in the high-frequency estimation with scaling arguments when the mean of the data generating process is different from zero.

\section{Methodology}
\label{sec:met}
\begin{figure}[h]
	\centering
	\includegraphics[width=\linewidth]{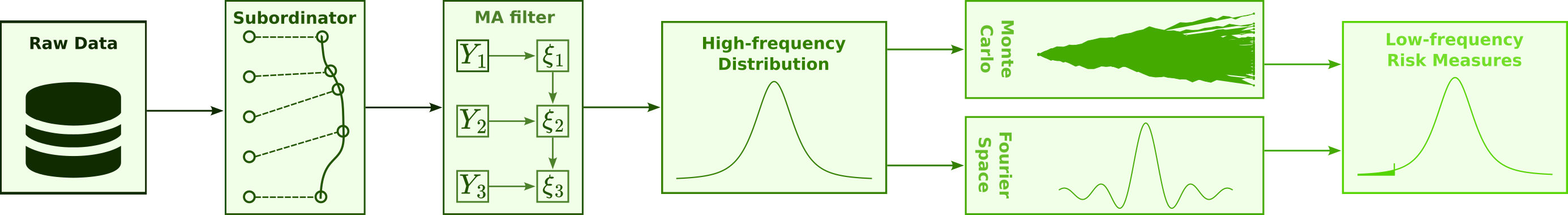}
	\caption{Scheme of the proposed approach.  First, the subordinator is applied to the raw data, and the subordinated returns time series is obtained. Optionally, the MA filter could be applied at this point to extract the innovations series. Then, a fat-tail distribution is fitted to the high-frequency data. Lastly, the low-frequency distribution is obtained via two parallel approaches based on Monte-Carlo simulation and characteristic function, and the daily risk measures are recovered by averaging them.}
	\label{img:graph_abstract}
\end{figure}

As stated in the previous Section, our working pipeline is similar to the one of  Ref. \cite{dimitriadis2022realized} and is schematized in Figure \ref{img:graph_abstract}. First, we apply a subordinator to the minute-by-minute data. Then, we apply a Moving Average (MA) filter on the data and fit a distribution on the residuals. Notice that the MA filter is optional, particularly if the data display non-significant auto-correlation. Finally, we apply a scaling procedure to estimate the risk measure at low frequency. It is worth mentioning that our approach is still in line with the two-step method proposed by \cite{clements2008quantile} (see the literature review above). The difference lies in the estimation of volatility, which is now implicit in the distributional assumption.\\ 
\\

As we show below in the result section,  this procedure significantly improves the benchmark of Ref. \cite{dimitriadis2022realized}, taking into account some key aspects of empirical asset prices.
Specifically, we move away from the assumption of a self-similar subordinated process, which is not supported by empirical evidence. In this way, we are able to provide a flexible framework for handling short-range autocorrelation in the data. Clearly, our procedure requires the choice, analytical or computational, of the distribution of subordinated returns (or residuals), leading to possible misspecifications. However, both simulations and empirical results indicate that a distributional misspecification, if any, affects the estimation of the risk measures less than the assumption of self-similarity with $H=1/2$. Finally, we recognize the estimation bias associated with a non-zero empirical mean and reduce its impact by assuming a fixed value for the location parameter of the distribution. The details are discussed in the following, along with the evaluation criteria used to compare the proposals.
\subsection{Subordinators}

The subordinator plays a key role in estimating realized risk measures and is defined as an injective function $\tau:[0,c] \rightarrow [0,390]$, with $\tau(0)=0$, $\tau(c)=390$, and $c\in\mathds{N}$ reflecting the desired number of intra-day returns, thus $c\le390$. It reflects the market activity by aggregating the high-frequency data over a coarse-grained grid in such a way that the activity of the market is the same in each interval. More precisely, let us define a non-negative process $\{\lambda_i\}_{i=0}^{390}$, also known as the intensity process, representing a quantity related to the considered asset. For example, $\lambda_i$ can be the traded volume in interval $i$ or a volatility measure. Then, we consider the cumulative sum $\{\Lambda_j\}_{j=0}^{390}$, with $\Lambda_j = \sum_{i=0}^j \lambda_i$ and the total sum $\Lambda=\sum_{i=0}^{390} \lambda_i$. The indexes set is thus partitioned in subsets $L_j:=\{ l\in\{1,\cdots,390\} | \frac{\Lambda}{c}(j-1) < \Lambda_l \le \frac{\Lambda}{c}j \}$ for $j=1,\cdots,c$ and, for each $j$, a $\tau(j)\in L_j$ is selected according to a specific criterion, such as $\tau(j) = \max L_j$. When working with real-world data, it can happen that some $L_j=\emptyset$. In such a case, $\tau(j)$ is drawn from a neighbor set, according to some tie-break rule (e.g., $\min L_{j+1}$).\\

Specifically, in the following, we consider three subordinators which are used independently in each experiment.
\begin{itemize}
    \item \textbf{Clock} is for the naive subordinator based on physical time, that is $\tau(j) = j \frac{390}{c}$, which corresponds to $\lambda_i=1 \ \forall i$.
    \item \textbf{TPV} is based on the Tri-Power Variation (see \cite{dong2017business} for further references) and captures the integrated variance. Specifically, we have used a centered window of 15 minutes to compute the TPV in a point $i$. That is:
    \begin{equation}
        \lambda_i = \sum_{l=\max(i-15, 0)+3} ^ {\min(i+15, 390)} (S_{l-3}-S_{l-2})^{2/3} (S_{l-2}-S_{l-1})^{2/3} (S_{l-1}-S_l)^{2/3}
    \end{equation}
    \item \textbf{Vol} uses the transaction volume as the intensity process.
\end{itemize}

\begin{figure}[h]
	\centering
	\includegraphics[scale=0.45]{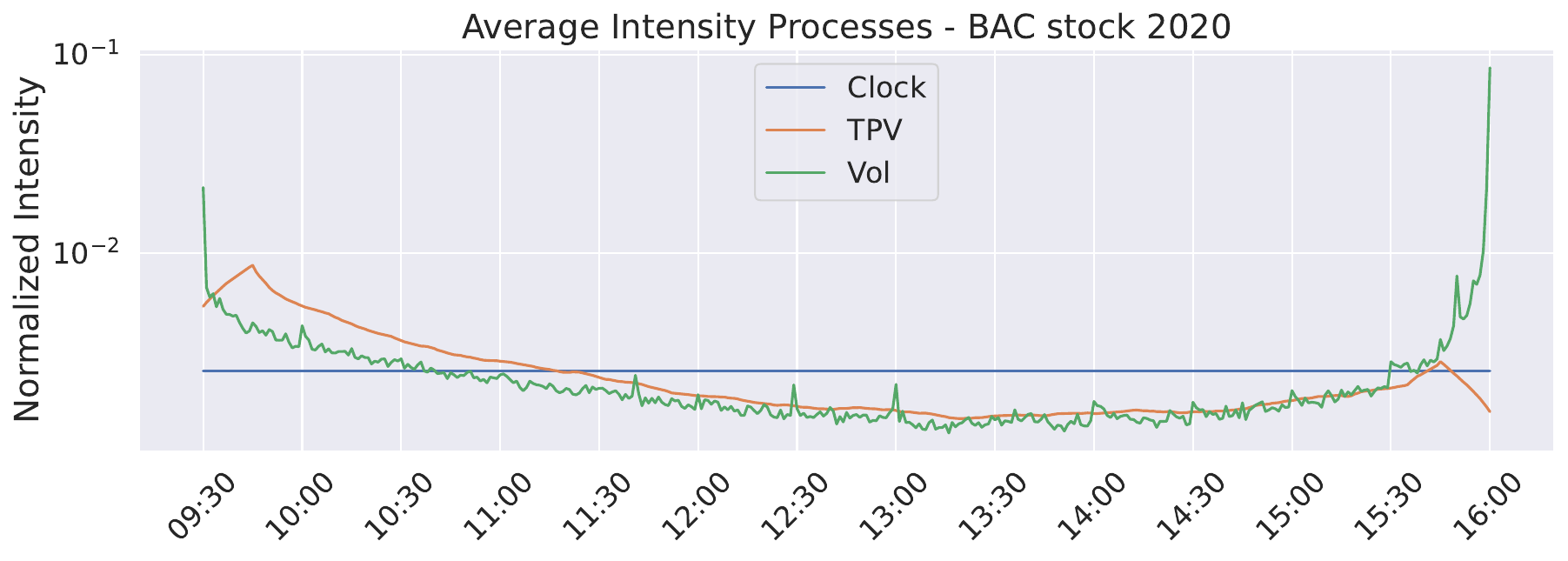}
	\caption{Comparison of the estimated intensity processes of different subordinators for the BAC stock in 2020.  The lines show the mean value of the daily intensity (normalized to sum up to 1).}
	\label{img:subs_comp}
\end{figure}

To provide an intuition of the subordinators' behavior, Figure \ref{img:subs_comp} shows the different intensity processes empirically measured on the BAC stock in 2020 (the daily intensity is averaged). Obviously, the Clock intensity is just a horizontal line, while Vol recovers the traditional asymmetric U-shape intraday pattern that characterizes trading volume in financial markets. It also displays some little peaks near round hours (e.g., 12:00, 12:30). Finally, the TPV intensity is larger in the first hours of the day, according to the known intraday pattern of volatility, and it is much smoother than the Vol intensity.
\subsection{Moving Average Filter}
As mentioned above, the Ljung-Box test on subordinated returns often (far more frequently than the test confidence level) rejects the null hypothesis of the absence of autocorrelation. To account for this effect, we model the $\{Y_j\}_{j=1}^c$ data as an MA(1) process. This choice is easy to handle, and we find that it captures most of the linear autocorrelation structure of subordinated returns. We estimate the parameters of the model
\begin{equation}
    \label{eq:1203_1100}
    Y_j = \phi \xi_{j-1} + \xi_j \quad \forall j=1,\cdots,c \ , \quad with \ \xi_j \ iid 
\end{equation}
via maximum likelihood, once a distributional assumption is made on $\xi_j$. The MA filter does not significantly increase the computational time required for the scaling to low-frequency, because, as we show below, we can analytically rewrite the daily characteristic function to simplify the computation. In the Monte Carlo approach, applying the filter is computationally equivalent to an element-wise sum and product. Thus, the main computational load is still in the sampling from the high-frequency distribution. Further details are provided in the remainder of this Section and in Appendix \ref{app_imp_det}.

\subsection{From High-frequency Data to Low-frequency Risk Measures}
\label{subsec:h2l}
As discussed previously, empirical prices are not well described by a monofractal process in subordinated time (see also \cite{cont2001empirical} and \cite{chakraborti2011econophysics}). For this reason, we model high-frequency log-price increments in subordinated time with a parametric family of fat-tailed distributions. Specifically, we choose a Student's t distribution, and we use maximum likelihood for parameter estimation.
It is important to emphasize the role of the distribution's mean $\mu$, which is traditionally estimated from the data as the sample mean. In our framework, this results in $\mu=\frac{y}{c}$, where $y=\sum_j y_j$ is the daily return. However, due to the fact that we scale the distribution to low-frequency before computing the risk measures, relying on the empirical mean would imply evaluating the risk measures of a distribution centered on the observed daily return $y$\footnote{Indeed, measuring the process at finer granularities does not give additional information or refinements on the mean estimate. To see this, let's assume the log-price process is an Arithmetic Brownian Motion $dS_t = \mu dt + \sigma dW_t$, where $W_t$ is a Brownian Motion and the time is measured on the daily time scale. If we sample $c+1$ equispaced observations $S_j$, then the $c$ increments are iid normally distributed with mean $\frac{\mu}{c}$ and standard deviation $\frac{\sigma}{\sqrt{c}}$. The empirical mean estimator is normally distributed with mean $\frac{\mu}{c}$ and standard deviation $\frac{\sigma}{c}$. When scaling to low-frequency, the estimated mean is multiplied by $c$, thus resulting in a normal distribution with mean $\mu$ and standard deviation $\sigma$. Thus, the signal-to-noise ratio is $\frac{\mu}{\sigma}$, which is independent on the number of sampled intra-day observations $c$.}.
An {\it a priori} assumption on $\mu$ could mitigate this issue. In particular, assuming zero mean could be reasonable for high-frequency data and is motivated by methodological simplicity and for a fair comparison with the method proposed by \cite{dimitriadis2022realized}, which implicitly assumes zero mean. However, neglecting the maximum likelihood estimation error comes at the cost of a possible bias error. Another possibility is to estimate $\mu$ as a long-term mean, e.g., the weekly/monthly average or using an Exponential Moving Average (EMA) scheme\footnote{From a conceptual point of view, the drawback of this solution refers to the use of historical daily observations, which are used together with high-frequency observations of one single day as input of the methodology.}. Given the time series of daily returns $\{y^{(t)}\}_{t=1,\cdots,T}$, the drift for the day $t$ is replaced by $EMA_t(\beta)$, which is defined as:
\begin{equation}
    EMA_{t}(\beta) = \frac{2}{\beta + 1} \cdot y^{(t-1)} + \left(1 - \frac{2}{\beta + 1}\right) \cdot EMA_{t-1}(\beta)
\end{equation}
The parameter $\beta$ controls the window length. The sample mean over one year of data is used to initialize $EMA_0(\beta)$.\\

After fitting the high-frequency distribution parameters, we find that the main concern of moving away from the Gaussian assumption is that the scaling to the low-frequency becomes more complex, as the closed-form for the sum of observations sampled from a general distribution is not known in closed form. This obstacle might be overcome in two distinct ways, one based on the characteristic function and one on a Monte Carlo simulation procedure.

To work with the characteristic function approach, we move in Fourier space. When the MA filter is not applied and the subordinated returns are modeled as iid random variables with Probability Density Function (PDF) $f_{HF}$ and characteristic function $\varphi_{HF}(\omega) := \mathds{E}\left[ \exp\{i\omega Y_j\} \right] = \int_{\mathds{R}} \exp\{i\omega x\} f_{HF}(x) dx$, the characteristic function of the daily return $Y$, $\varphi_Y$, can be easily computed as:
\begin{equation}
    \varphi_Y(\omega) = \mathds{E}\left[ \exp\{i\omega Y\} \right] = \mathds{E}\left[ \exp\left\{i\omega \sum_{j=1}^c Y_j \right\} \right] = \prod_{j=1}^c \mathds{E}\left[ \exp\{i\omega Y_j\} \right] = \left[\varphi_{HF}(\omega)\right]^c
\end{equation}

When using the MA filter, we indicate with $\varphi_{HF}$ the characteristic function of the innovations $\xi_j$ in Eq. \eqref{eq:1203_1100} and $\varphi_Y$ reads:
\begin{equation}
    \varphi_Y(\omega) = \varphi_{HF}(\omega) \cdot \varphi_{HF}(\phi\omega) \cdot \left[\varphi_{HF}\left( (1+\phi)\omega \right)\right]^{c-1}
\end{equation}
The characteristic function $\varphi_Y$ can be used to recover the CDF $F_Y$, thanks to the Gil-Pelaez formula \cite{gil1951note}:
\begin{equation}
    \label{eq:1217_1340}
    F_{Y}(x) = \frac{1}{2} - \frac{1}{\pi}\int_0^\infty Im\left( \frac{\exp(-i\omega x) \varphi_{Y}(\omega)}{\omega}  \right) d\omega
\end{equation}
where $Im(\cdot)$ is the imaginary part. Finally, VaR and ES are recovered from $F_Y$ with a customized numerical routine mainly based on Brent's method \cite{brent2013algorithms} method, see Appendix \ref{app_imp_det} for further details. Algorithm \ref{algo:ch_fun} shows the pseudo-code of the characteristic function approach in the iid case. Here \texttt{quad} denotes a quadrature algorithm taking as input the integrated function, the integration variable, and the domain. The routine \texttt{zero\_search} is used for finding the zero of the function given as input. This approach is not prone to sampling errors as no randomness is involved in the computation, and working in a controlled environment could provide a good basis for investigating the analytical properties of the estimator. However, it relies on the numerical approximations of the daily CDF and risk measures, which might introduce numerical errors and make the computation heavy.\\

\begin{algorithm}[h]
\caption{Characteristic function approach for risk scaling - iid case}
\label{algo:ch_fun}
\begin{algorithmic}[1] 
\State \textbf{Input}: $\varphi_{HF}$, $c$, $\theta$, \texttt{quad}, \texttt{zero\_search}
\State \textbf{Output}: \texttt{q\_hat}, \texttt{e\_hat}.
\State \textbf{define function} \texttt{GP\_Integrand}($\omega$, $x$, $\varphi_{HF}$, $c$):
\State $\quad$ Compute \texttt{argument} = \texttt{exp}$(-i*\omega * x)*$\texttt{power}$(\varphi_{HF}(\omega), c)$.
\State $\quad$ \textbf{return} \texttt{argument.imag}
\State \textbf{define function} \texttt{Target\_Opt}($x$, $\theta$, $\varphi_{HF}$, $c$):
\State $\quad$ Compute \texttt{integral} = \texttt{quad}(\texttt{GP\_Integrand}($\omega$, $x$, $\varphi_{HF}$, $c$), $\omega\in[0,\infty)$)
\State $\quad$ \textbf{return} 0.5 - \texttt{integral} / $\pi$ - $\theta$
\State Initialize \texttt{q\_list = list()}
\For{$j \gets 1 \text{ to } 10$}
    \State Compute $\theta_j = \frac{j}{10}\theta$.
    \State Compute \texttt{q\_temp} = \texttt{zero\_search}(\texttt{Target\_Opt}($x$, $\theta_j$, $\varphi_{HF}$, $c$))
    \State Append \texttt{q\_temp} to \texttt{q\_list}
\EndFor
\State Define \texttt{q\_hat = q\_list[-1]}
\State Define \texttt{e\_hat = q\_list.mean()}
\end{algorithmic}
\end{algorithm}

The alternative approach is to approximate the daily return distribution via Monte-Carlo simulations.  In the iid case, we randomly draw a $B\times c$ matrix $\{z_{b,j}\}_{b=1\cdots B}^{j=1,\cdots,c}$, where $B\in\mathds{N}$ represent the batch size and each entry $z_{b,j}$ is distributed as the high-frequency observations. Then, we sum the elements in each row, and the output vector of size $B$ is used as a proxy for the $Y$ distribution. Finally, VaR and ES are numerically computed from these simulated values. When using the MA filter, we instead simulate the innovations $\xi_j$. Algorithm \ref{algo:mc} describes in a schematic way the procedure for the iid case. Specifically, \texttt{random\_hf} is a random generator for the high-frequency distribution, and \texttt{quantile} is a routine for computing the empirical quantile. This approach is prone to sampling approximation errors as the Monte-Carlo simulation replaces the true $Y$ distribution. We reduce the numerical errors as simpler operations are involved and the computational time is acceptable as the simulation runs in less than half a second on a standard laptop.\\

\begin{algorithm}[h]
\caption{Monte-Carlo approach for risk scaling  - iid case}
\label{algo:mc}
\begin{algorithmic}[1] 
\State \textbf{Input}: $c$, $\theta$, $B$, \texttt{random\_hf}, \texttt{quantile}
\State \textbf{Output}: \texttt{q\_hat}, \texttt{e\_hat}.
\State Initialize \texttt{daily\_dist = list()}
\For{$b \gets 1 \text{ to } B$}
    \State Initialize \texttt{daily\_ret = 0}
    \For{$j \gets 1 \text{ to } c$}
        \State Update \texttt{daily\_ret += random\_hf()}
    \EndFor
    \State Append \texttt{daily\_ret} to \texttt{daily\_dist}
\EndFor
\State Compute \texttt{q\_hat = quantile}(\texttt{daily\_dist}, $\theta$).
\State Compute \texttt{e\_hat = daily\_dist[daily\_dist < q\_hat].mean()}
\end{algorithmic}
\end{algorithm}

To summarize, as both the aggregation procedures described have some weaknesses, in the following we average their output. on order to obtain a more robust estimator.
\subsection{Evaluate and Compare Estimation Approaches}
\label{subsec:ev_comp}
Backtesting realized risk measures estimated with different approaches is not trivial. To see this, let us focus on two timesteps, which are the opening and the closing of the regular trading hour.
The main issue is that the estimation approaches use information from the whole day, which is the information set available at the closing time.

Generally speaking, this could introduce a bias that makes the evaluation step troublesome. One would be tempted to use the Pinball and ES losses $\mathcal{L}_q^\theta(\hat{q}, Y)$ and $\mathcal{L}_e^\theta(\hat{q}, \hat{e}, Y)$ defined in Eqs. \eqref{eq:1113_0900} and \eqref{eq:1113_0901}, as we know that their conditional expectation is minimized by the conditional VaR/ES. The problem is that we are conditioning on $\mathcal{F}_e$, which is the information set we are using to find $\hat{q}$ and $\hat{e}$. Thus, $Y$ is measurable and the argmin is attained in $Y$. This is inconsistent with the fact that, almost surely, $VaR_\theta(Y)\neq Y$ and $ES_\theta(Y)\neq Y$. In other words, as $Y|\mathcal{F}_e$ is a degenerate random variable, we can take the calculus perspective by regarding $Y$ as a deterministic variable and searching for the argmin of the losses. It coincides with $Y$, and the loss monotonically increases as we move away from the observed value. Generally, this is not a weakness for the usual forecasting setting, as $Y$ is not observable when the prediction is made. But in our framework, $Y$ is measurable. So, using the losses as a benchmark for evaluating the goodness of the proposed methodologies leads to incoherent results, as reported in more detail in Appendix \ref{app:loss_bias}.\\

Following the previous reasoning, we evaluate the estimation approaches using three alternative strategies. Let us consider a dataset made up of several days, and let $\{y^{(t)}\}_{t=1,\cdots,T}$, $\{\hat{q}^{(t)}\}_{t=1,\cdots,T}$, and $\{\hat{e}^{(t)}\}_{t=1,\cdots,T}$ be the time series of daily observations and estimated risk measures. The first strategy is to use the statistical properties to understand whether the measures effectively describe the tail. In this way, we can perform an in-sample evaluation. As for the VaR, a common choice consists of counting the hits frequency, i.e. the percentage of observed values below the quantile, that is $\frac{1}{T} \sum_{t=1}^T \mathds{1}_{\{y^{(t)} \le \hat{q}^{(t)}\}}$. If the estimator is correctly specified, the frequency converges to the probability level $\theta$ as the sample size increases. Regarding the ES, we use bootstrap tests based on the $Z_1$ and $Z_2$ statistics proposed in \cite{acerbi2014back}. The underlying idea is to rewrite the ES definition to obtain the target value (under the null hypothesis of correctly specified VaR and ES) for some specific statistics. Specifically, the $Z_1$ statistic is defined as $Z_1 := \frac{Y}{ES_\theta(Y)}$. The target value is obtained as:
\begin{equation}
    ES_\theta(Y) = \mathds{E}[ Y | Y \le VaR_\theta(Y)] \implies \mathds{E}[ Z_1 | Y \le VaR_\theta(Y) ] = 1
\end{equation}
Instead, the $Z_2$ statistic is $Z_2 := \frac{Y}{\theta ES_\theta(Y)} \mathbbm{1}_{\{Y\le VaR_\theta(Y)\}}$, whose target value is:
\begin{equation}
    ES_\theta(Y) = \mathds{E}[ Y | Y \le VaR_\theta(Y)] = \frac{1}{\theta} \mathds{E}[ Y \mathbbm{1}_{\{Y\le VaR_\theta(Y)\}}] \implies \mathds{E}[ Z_2 ] = 1
\end{equation}
In the following, we indicate the bootstrap tests based on the $Z_1$ and $Z_2$ statistics as AS1 and AS2, respectively.\\

An alternative evaluation strategy, adopted in Ref. \cite{dimitriadis2022realized}, consists of using the realized measures out-of-sample. Here, we are forced to make assumptions about the dynamical pattern of the risk measures over several days. We choose to work in a simple setting and we assume that the time series $\{\hat{q}^{(t)}\}_{t=1,\cdots,T}$ and $\{\hat{e}^{(t)}\}_{t=1,\cdots,T}$ follow an AR process\footnote{Further experiments with the Random Walk model and the Exponential Moving Average have been performed and shown in Appendix \ref{app:for_rrm}.}. Thus, we first fit the AR model on a subset of the two time series $\{\hat{q}^{(t)}\}_{t=1,\cdots,T}$ and $\{\hat{e}^{(t)}\}_{t=1,\cdots,T}$ used as the train set. Later, the out-of-sample forecasts are obtained, and they are evaluated using the losses $\mathcal{L}_q^\theta(\hat{q}, Y)$ and $\mathcal{L}_e^\theta(\hat{q}, \hat{e}, Y)$. In this way, we are not using $\hat{q}^{(t+1)}$ and $\hat{e}^{(t+1)}$ as an estimate for the risk at time $t+1$. Rather, we use their forecast obtained with the information up to time $t$ and thus we do not face the incoherence previously mentioned.\\

Finally, the third approach, which can be used only when working with simulated data, consists of measuring the distance from the ground truth. This can be straightforwardly computed as the $l_2$ norm of the risk measures.
\section{Numerical and Empirical Results}
\label{sec:exp}
In this section, we present the experimental results. First, we consider numerical simulations of data generating processes for which we know the ground truth value of the risk measures. With this approach, it is possible to gain a deeper understanding of the finite-sample properties of the proposed estimators. Next, we investigate two real-world datasets and the estimation approaches are compared using the in-sample tests and the out-of-sample predictive accuracy previously described. The overall results show that our framework provides the best results. 

Despite the different considered datasets, the experimental scheme is the same in all cases. We test both the approach in \cite{dimitriadis2022realized} (from now on identified with \textbf{DH}) and ours with the Student's t distribution to model the returns or the residuals. We make this choice because it is simple to handle and popular in financial applications. Nevertheless, we again stress that our framework is completely general and that any other distribution can be used. Next, we compare the three subordinators previously described. Moreover, we consider both the setting with the MA filter and the one without it, either with the drift specification $\mu=0$ (\textbf{t-iid} and \textbf{t-MA}) or with EMA(21) (\textbf{t-iid (21)} and \textbf{t-MA (21)}) -i.e., the Exponential Moving Average with monthly window length. Then, we compare three different numbers of intra-day returns, namely $c=39,78,130$. As observed before, when using the Clock subordinator, they correspond to 3, 5, and 10-minute frequencies. We work with multiple $c$ values to explore the well-known bias-variance trade-off associated with sampling frequency. Larger values of $c$ provides more accurate estimation of the high-frequency distribution because of the larger sample. However, larger $c$ are associated with larger (microstructural) noise for the intra-day returns.
Regarding the target probability level for the risk measure, we consider $\theta=0.05, 0.025, 0.01$ for the VaR estimation and $\theta=0.05, 0.025$ for the ES, as standard in the literature.\\

Regarding the evaluation, a rolling cross-validation is performed. In more detail, we have already specified that every day is considered an independent episode. However, in the evaluation step, there is a need for aggregation. Specifically, the ES tests require a population of more days. In this case, we adopt a rolling-fold strategy, and one test is performed for every year of data. Moreover, also the out-of-sample evaluation requires aggregation of more days to obtain a train and a test set. In this case, in line with \cite{gatta2024caesar}, we exploit block rolling windows of 5 years for the train and one for the test. Lastly, to present the results, we aggregate them across the different assets and rolling windows, and we display the mean values.
\subsection{Synthetic Datasets}
\label{subsec:synth}
The first study is performed on a synthetic dataset replicating the subordinated returns. In this way, there is no need to use the subordinators, and the different approaches are directly compared. The dataset is made up of twelve time series spanning ten years of data each. There are four time series for each $c=39, 78, 130$ value, each generated using a distinct data-generating process. The intra-day data $\{Y_j\}_{j=1}^c$ are sampled according to an iid or an MA(1) process $Y_j = \phi \xi_{j-1} + \xi_j$, where the observations $\{Y_j\}_{j=1}^c$ (iid case) or the innovations $\{\xi_j\}_{j=0}^c$ (MA case) are sampled either from a Gaussian or a Student's t distribution.
In all cases, the model parameters are set equal to the best fit of the real-world banking dataset; see Section \ref{sec:dataset} for further details. Specifically, for each day, we consider the TPV subordinator and the empirical return distributions is fitted. The median parameter values obtained from this procedure are then used to define the parameters of the data-generating processes. We end up with twelve sets of parameters, one for every generated time series. Roughly, $\phi$ values in the MA series are close to -0.05, and the degrees of freedom $\nu$ of the Student's t series are between 2 and 4. Instead, the location and scaling parameters $\mu$ and $\sigma$ significantly vary with $c$.\\

The comparison between different approaches is made using the root mean squared error (rMSE) between the estimated risk measure and the true one. In the Gaussian case, we know that:
\begin{equation}
    Y = \sum_{j=1}^c Y_j = \sum_{j=1}^c (\phi \xi_{j-1} + \xi_j) \implies Y = \phi \xi_0 + \sum_{j=1}^{c-1} (1+\phi)\xi_j + \xi_c
\end{equation}
As such, the daily return is still normally distributed with mean $c(1+\phi)\mu$ and variance $[(c-1)(1+\phi)^2 + (1+\phi^2)]\sigma^2$. Then, the VaR and ES can be analytically computed; see \cite{nadarajah2014estimation}. As for the Student's t case, there is no analytical form for the sum, so we rely on Monte-Carlo simulation to map the intra-day coefficients into the daily distribution and on empirical estimators to extract its risk measures. The rMSE for the VaR and the ES are displayed in Table \ref{tab:sim_q_distance} and  Table \ref{tab:sim_e_distance}, respectively. Instead, the in-sample tests (for a comparison with the real-world dataset) are shown in Appendix \ref{app:loss_bias}, Tables \ref{tab:sim_q_intests} and \ref{tab:sim_e_intests}. To corroborate the discussion in Section \ref{subsec:ev_comp}, in the same appendix, we also show the in-sample $\mathcal{L}_q^\theta$ and $\mathcal{L}_e^\theta$ losses in Table \ref{tab:sim_q_loss} and \ref{tab:sim_e_loss}, respectively.
\begin{table}[h]
\begin{center}
{\small

\begin{tabular}{lccccccccc}
\toprule
\multicolumn{10}{c}{\textbf{GAUSSIAN SYNTHETIC DATASET}} \\
\toprule
\multirow{2}{*}{\textbf{Algorithm}} & \multicolumn{3}{c}{$\theta=0.05$} & \multicolumn{3}{c}{$\theta=0.025$} & \multicolumn{3}{c}{$\theta=0.01$} \\
\cmidrule(lr){2-4} \cmidrule(lr){5-7} \cmidrule(lr){8-10}
 & $c=39$ & $c=78$ & $c=130$ & $c=39$ & $c=78$ & $c=130$ & $c=39$ & $c=78$ & $c=130$ \\
\midrule
\textbf{DH} & 4.005 & 3.004 & 2.385 & 4.894 & 3.722 & 2.966 & 6.183 & 4.690 & 3.825\\
\textbf{t-iid} & \textbf{2.544} & \textbf{1.893} & \textbf{1.587} & \textbf{3.053} & \textbf{2.283} & \textbf{1.920} & \textbf{3.700} & \textbf{2.773} & \textbf{2.276}\\
\textbf{t-MA} & \underline{2.876} & \underline{2.305} & \underline{2.003} & \underline{3.458} & \underline{2.742} & \underline{2.393} & \underline{4.220} & \underline{3.267} & \underline{2.844}\\
\textbf{t-iid (5)} & 6.034 & 5.974 & 6.082 & 6.275 & 6.115 & 6.185 & 6.630 & 6.320 & 6.309\\
\textbf{t-MA (5)} & 6.183 & 6.084 & 6.179 & 6.486 & 6.263 & 6.316 & 6.944 & 6.514 & 6.504\\
\textbf{t-iid (21)} & 3.629 & 3.336 & 3.271 & 4.009 & 3.565 & 3.454 & 4.535 & 3.867 & 3.668\\
\textbf{t-MA (21)} & 3.865 & 3.548 & 3.457 & 4.321 & 3.841 & 3.694 & 4.964 & 4.231 & 4.007\\
\bottomrule
\end{tabular}

\vspace{0.1cm}

\begin{tabular}{lccccccccc}
\toprule
\multicolumn{10}{c}{\textbf{STUDENT'S T SYNTHETIC DATASET}} \\
\toprule
\multirow{2}{*}{\textbf{Algorithm}} & \multicolumn{3}{c}{$\theta=0.05$} & \multicolumn{3}{c}{$\theta=0.025$} & \multicolumn{3}{c}{$\theta=0.01$} \\
\cmidrule(lr){2-4} \cmidrule(lr){5-7} \cmidrule(lr){8-10}
 & $c=39$ & $c=78$ & $c=130$ & $c=39$ & $c=78$ & $c=130$ & $c=39$ & $c=78$ & $c=130$ \\
\midrule
\textbf{DH} & 15.143 & 15.882 & 16.964 & 22.020 & 19.773 & 19.384 & 55.899 & 42.034 & 29.740\\
\textbf{t-iid} & \textbf{9.840} & \textbf{9.126} & \textbf{8.497} & \underline{13.784} & \textbf{12.585} & \textbf{11.743} & 21.990 & 19.988 & \textbf{18.759}\\
\textbf{t-MA} & \underline{9.923} & \underline{9.344} & \underline{8.843} & \textbf{13.725} & \underline{12.710} & \underline{12.062} & \textbf{21.448} & \textbf{19.823} & 18.937\\
\textbf{t-iid (5)} & 14.855 & 15.022 & 14.486 & 17.664 & 17.297 & 16.457 & 24.542 & 23.196 & 21.819\\
\textbf{t-MA (5)} & 14.764 & 15.012 & 14.560 & 17.456 & 17.221 & 16.521 & 23.888 & 22.862 & 21.768\\
\textbf{t-iid (21)} & 11.041 & 10.214 & 9.587 & 14.549 & 13.207 & 12.324 & 22.316 & 20.126 & \underline{18.808}\\
\textbf{t-MA (21)} & 10.990 & 10.262 & 9.780 & 14.382 & 13.176 & 12.512 & \underline{21.695} & \underline{19.832} & 18.867\\
\bottomrule
\end{tabular}

\caption{Synthetic dataset - root Mean Squared Error ($\times 10^3$) between the estimated VaR and the ground truth. The best result is in bold, and the second-best is underlined.}
\label{tab:sim_q_distance}
}
\end{center}
\end{table}

\begin{table}[h]
\begin{center}
\begin{adjustwidth}{-0.9cm}{}
{\small

\begin{tabular}{lcccccc||cccccc}
\toprule
 & \multicolumn{6}{c}{\textbf{GAUSSIAN SYNTHETIC DATASET}} & \multicolumn{6}{c}{\textbf{STUDENT'S T SYNTHETIC DATASET}} \\
\cmidrule(lr){2-7} \cmidrule(lr){8-13}
\multirow{2}{*}{\textbf{Algorithm}} & \multicolumn{3}{c}{$\theta=0.05$} & \multicolumn{3}{c}{$\theta=0.025$} & \multicolumn{3}{c}{$\theta=0.05$} & \multicolumn{3}{c}{$\theta=0.025$} \\
\cmidrule(lr){2-4} \cmidrule(lr){5-7} \cmidrule(lr){8-10} \cmidrule(lr){11-13}
 & $c=39$ & $c=78$ & $c=130$ & $c=39$ & $c=78$ & $c=130$ & $c=39$ & $c=78$ & $c=130$ & $c=39$ & $c=78$ & $c=130$ \\
\midrule
\textbf{DH} & 0.499 & 0.365 & 0.283 & 0.648 & 0.470 & 0.359 & 4.690 & 3.917 & 3.113 & 8.325 & 6.951 & 4.857\\
\textbf{t-iid} & \textbf{0.319} & \textbf{0.239} & \textbf{0.197} & \textbf{0.369} & \textbf{0.279} & \textbf{0.225} & 2.036 & 1.973 & 1.936 & 2.920 & 2.836 & 2.800\\
\textbf{t-MA} & \underline{0.365} & \underline{0.282} & \underline{0.245} & \underline{0.427} & \underline{0.322} & \underline{0.279} & \textbf{1.992} & \underline{1.913} & \underline{1.861} & \textbf{2.839} & \underline{2.737} & \underline{2.680}\\
\textbf{t-iid (5)} & 0.634 & 0.615 & 0.620 & 0.662 & 0.632 & 0.630 & 2.294 & 2.266 & 2.205 & 3.099 & 3.037 & 2.973\\
\textbf{t-MA (5)} & 0.659 & 0.630 & 0.634 & 0.697 & 0.649 & 0.648 & 2.250 & 2.218 & 2.142 & 3.017 & 2.948 & 2.862\\
\textbf{t-iid (21)} & 0.412 & 0.361 & 0.347 & 0.453 & 0.385 & 0.364 & 2.060 & 1.962 & 1.914 & 2.920 & 2.801 & 2.752\\
\textbf{t-MA (21)} & 0.448 & 0.390 & 0.373 & 0.501 & 0.419 & 0.396 & \underline{2.017} & \textbf{1.908} & \textbf{1.847} & \underline{2.840} & \textbf{2.707} & \textbf{2.639}\\
\bottomrule
\end{tabular}
}
\end{adjustwidth}

\caption{Synthetic dataset - root Mean Squared Error ($\times 10^2$) between the estimated ES and the ground truth.}
\label{tab:sim_e_distance}
\end{center}
\end{table}

The tables show the rMSE of the realized measures obtained with the characteristic function approach and the Monte-Carlo simulation, as well as their average. In general, the proposed approach outperforms the method by \cite{dimitriadis2022realized}. In the Student's t dataset, the MA version outperforms the other, as expected. Interestingly, the estimates that do not consider the MA filter perform better in the Gaussian dataset, that is, when the shape assumption is violated. This result may appear counterintuitive, given that the data-generating process is itself an MA model. However, the key lies in the relatively small value of the autocorrelation coefficient obtained from the real-data fit. Indeed, when the coefficient is increased ($\phi=-0.2$), then the MA-based version of the model outperforms the iid specification, even with Gaussian distributed innovations. As for the performance behaviour in relation to the probability level $\theta$, we observe a general degradation as $\theta$ becomes smaller, which is consistent with the "deep-tail" effect observed in \cite{gatta2024caesar}. Moreover, there is an improvement for larger values of $c$, due to the fact that there is no noise in the observations.\\


We also study the sensitivity of the methods to variations in the mean of the data-generating process. 
\begin{figure}[h]
	\centering
	\includegraphics[scale=0.40]{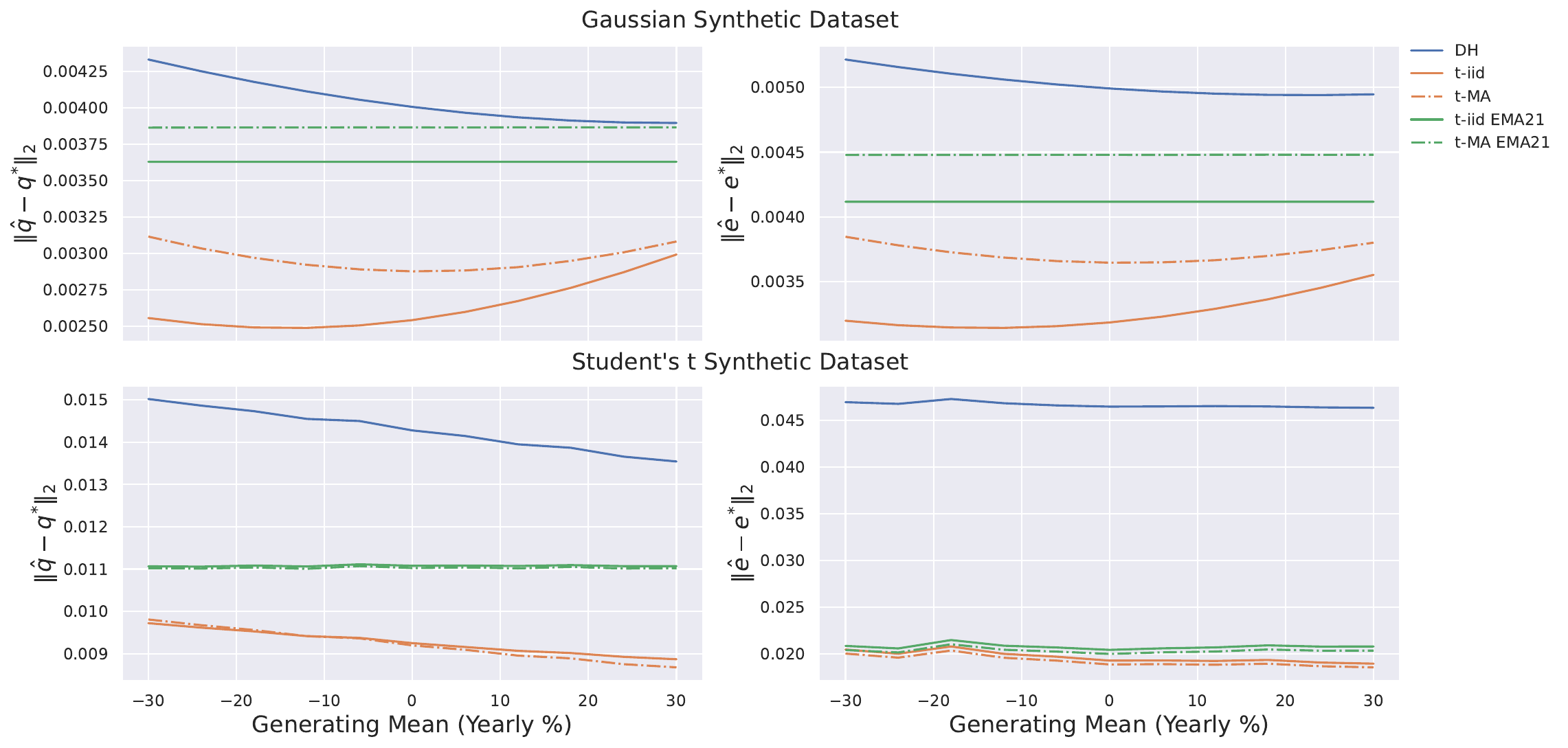}
	\caption{rMSE against the mean of the generative process.  The rMSE is computed as the distance between the estimated risk measure $\hat{q}$ (or $\hat{e}$) and the ground truth $q^{*}$ (or $e^{*}$), at the probability level $\theta=0.05$ and $c=39$. Five approaches are compared: \textbf{DH} (blue line), \textbf{t-iid} (orange line), \textbf{t-MA} (orange dotted line), \textbf{t-iid (21)} (green line), and \textbf{t-MA (21)} (green dotted line). The upper panels are for the Gaussian synthetic dataset, while the bottom ones are for the Student's t synthetic dataset. The left panels refer to the VaR, whereas the right panels refer to the ES. The mean process values on the x-axis are annualized.}
	\label{img:rmse_vs_mu_appendix_with_ema}
\end{figure}
As shown in Figure \ref{img:rmse_vs_mu_appendix_with_ema}, the performance of \textbf{t-iid} and \textbf{t-MA} our method for the Gaussian case exhibits a mild decline as the absolute value of the mean in the data-generating process increases. Nevertheless, it still outperforms for values that make sense within the financial context. This sensitivity is due to the use of zero as the location parameter in the distribution fitting step. The EMA-based approaches with parameter $\beta=21$ (lines \textbf{t-iid (21)} and \textbf{t-MA (21)}) overcome this issue\footnote{The lines in the plot as almost straight. This is due to the fact that a separate experiment is carried out for every mean value on the x-axis. The random seed in each experiment is initialized to the same value. Practically, this implies that the time series for different mean values are obtained by shifting the same one. Thus, after removing the mean using the EMA, the remaining part of our pipeline is actually executed on nearly the same series regardless of the underlying mean of the data-generating process. As a result, the associated error is almost constant. Clearly, even when varying the random seed, the differences across different mean values are minimal and can be considered negligible.}. However, in the synthetic dataset and with the $\theta$ and $c$ values considered, it performs worse than the zero-mean version.

\subsection{Empirical Datasets}
\label{sec:dataset}
The empirical analysis is performed on two datasets of equity prices:
\begin{itemize}
    \item \textbf{Bank} comprises eight stocks of the US banking sector, as in \cite{keilbar2022modelling}. Specifically, the stocks are Bank of America (BAC), Bank of New York Mellon (BK), Citigroup (C), Goldman Sachs (GS), JPMorgan (JPM), Morgan Stanley (MS), State Street Corp (STT), and Wells Fargo (WFC). The period spans from 1998 to 2020 (except for MS, starting from 2006, and GS, beginning in 2000).
    \item \textbf{Nasdaq} includes the 10 stocks in the Nasdaq index with large capitalization. Namely, they are Apple (AAPL), Amazon (AMZN), Broadcom (AVGO), Costco (COST), Meta (FB), Google (GOOGL), Microsoft (MSFT), Netflix (NFLX), Nvidia (NVDA), and Tesla (TSLA). Generally, the period considered is from 1998 to 2020, but for some stocks, less data are available.
    \end{itemize}
Both data sets are made up of minute-by-minute observations of the trade price, and only regular trading hours are considered. Any missing value is filled with the previous available price. 

The results on hits frequency are shown in Table \ref{tab:real_hits}.  The main result is that the proposed estimators with the $EMA(5)$ drift usually perform better than the other approaches and achieve frequencies very close to the target value $\theta$. Furthermore, our filters generally outperform \textbf{DH}. Moreover, the Clock and TPV subordinators outperform the one based on trading volume, but it is not possible to have a clear indication of what is the best. Furthermore, the performance of all the approaches seems to be slightly degraded as $c$ increases, differently from what happens in the synthetic dataset. As already pointed out, this is likely due to the presence of noise in the raw observations.\\

\begin{table}[H]
\begin{center}
{\small

\begin{tabular}{lccccccccc}
\toprule
\multicolumn{10}{c}{\textbf{BANK DATASET}} \\
\toprule
\multirow{2}{*}{\textbf{Algorithm}} & \multicolumn{3}{c}{$\theta=0.05$} & \multicolumn{3}{c}{$\theta=0.025$} & \multicolumn{3}{c}{$\theta=0.01$} \\
\cmidrule(lr){2-4} \cmidrule(lr){5-7} \cmidrule(lr){8-10}
 & $c=39$ & $c=78$ & $c=130$ & $c=39$ & $c=78$ & $c=130$ & $c=39$ & $c=78$ & $c=130$ \\
\midrule
\textbf{DH (Clock)} & 0.037 & 0.037 & 0.036 & 0.014 & 0.013 & 0.012 & 0.004 & 0.003 & 0.003\\
\textbf{DH (TPV)} & 0.035 & 0.034 & 0.032 & 0.018 & 0.015 & 0.013 & 0.008 & 0.005 & 0.004\\
\textbf{DH (Vol)} & 0.033 & 0.031 & 0.029 & 0.015 & 0.012 & 0.010 & 0.006 & 0.003 & 0.002\\
\midrule
\textbf{t-iid (Clock)} & 0.038 & 0.038 & 0.035 & 0.015 & 0.015 & 0.014 & 0.004 & 0.005 & 0.004\\
\textbf{t-iid (TPV)} & 0.041 & 0.038 & 0.030 & 0.017 & 0.016 & 0.013 & 0.005 & 0.004 & 0.004\\
\textbf{t-iid (Vol)} & 0.036 & 0.032 & 0.023 & 0.015 & 0.013 & 0.010 & 0.004 & 0.004 & 0.003\\
\midrule
\textbf{t-MA (Clock)} & 0.041 & 0.040 & 0.039 & 0.016 & 0.016 & 0.017 & 0.004 & 0.005 & 0.005\\
\textbf{t-MA (TPV)} & 0.043 & 0.041 & 0.033 & 0.017 & 0.018 & 0.014 & 0.005 & 0.005 & 0.005\\
\textbf{t-MA (Vol)} & 0.037 & 0.033 & 0.025 & 0.015 & 0.014 & 0.010 & 0.003 & 0.004 & 0.003\\
\midrule
\textbf{t-iid (5; Clock)} & 0.057 & 0.055 & \textbf{0.051} & \underline{0.026} & \textbf{0.025} & \textbf{0.024} & 0.009 & 0.009 & \underline{0.009}\\
\textbf{t-iid (5; TPV)} & 0.057 & \textbf{0.053} & 0.041 & 0.027 & \underline{0.026} & 0.020 & \textbf{0.009} & \underline{0.010} & 0.007\\
\textbf{t-iid (5; Vol)} & \textbf{0.052} & 0.044 & 0.030 & \textbf{0.024} & 0.021 & 0.014 & 0.008 & 0.007 & 0.005\\
\midrule
\textbf{t-MA (5; Clock)} & 0.060 & 0.058 & 0.056 & 0.027 & 0.028 & \underline{0.027} & 0.007 & 0.009 & \textbf{0.009}\\
\textbf{t-MA (5; TPV)} & 0.060 & 0.059 & \underline{0.046} & 0.029 & 0.028 & 0.022 & \underline{0.009} & \textbf{0.010} & 0.008\\
\textbf{t-MA (5; Vol)} & \underline{0.053} & \underline{0.045} & 0.032 & 0.024 & 0.022 & 0.015 & 0.007 & 0.007 & 0.005\\
\midrule
\textbf{t-iid (21; Clock)} & 0.042 & 0.041 & 0.039 & 0.018 & 0.017 & 0.017 & 0.005 & 0.005 & 0.005\\
\textbf{t-iid (21; TPV)} & 0.044 & 0.041 & 0.032 & 0.019 & 0.018 & 0.014 & 0.006 & 0.006 & 0.005\\
\textbf{t-iid (21; Vol)} & 0.039 & 0.034 & 0.024 & 0.017 & 0.015 & 0.010 & 0.005 & 0.004 & 0.003\\
\midrule
\textbf{t-MA (21; Clock)} & 0.045 & 0.044 & 0.043 & 0.018 & 0.019 & 0.018 & 0.004 & 0.005 & 0.006\\
\textbf{t-MA (21; TPV)} & 0.047 & 0.045 & 0.036 & 0.020 & 0.019 & 0.016 & 0.005 & 0.006 & 0.005\\
\textbf{t-MA (21; Vol)} & 0.040 & 0.036 & 0.026 & 0.016 & 0.015 & 0.011 & 0.004 & 0.005 & 0.003\\
\midrule
\bottomrule
\end{tabular}

\vspace{0.1cm}

\begin{tabular}{lccccccccc}
\toprule
\multicolumn{10}{c}{\textbf{NASDAQ DATASET}} \\
\toprule
\multirow{2}{*}{\textbf{Algorithm}} & \multicolumn{3}{c}{$\theta=0.05$} & \multicolumn{3}{c}{$\theta=0.025$} & \multicolumn{3}{c}{$\theta=0.01$} \\
\cmidrule(lr){2-4} \cmidrule(lr){5-7} \cmidrule(lr){8-10}
 & $c=39$ & $c=78$ & $c=130$ & $c=39$ & $c=78$ & $c=130$ & $c=39$ & $c=78$ & $c=130$ \\
\midrule
\textbf{DH (Clock)} & 0.031 & 0.030 & 0.028 & 0.012 & 0.008 & 0.008 & 0.004 & 0.002 & 0.002\\
\textbf{DH (TPV)} & 0.028 & 0.026 & 0.023 & 0.014 & 0.011 & 0.008 & 0.007 & 0.003 & 0.002\\
\textbf{DH (Vol)} & 0.026 & 0.024 & 0.022 & 0.012 & 0.009 & 0.006 & 0.005 & 0.003 & 0.001\\
\midrule
\textbf{t-iid (Clock)} & 0.034 & 0.032 & 0.029 & 0.012 & 0.012 & 0.011 & 0.003 & 0.003 & 0.003\\
\textbf{t-iid (TPV)} & 0.035 & 0.032 & 0.023 & 0.014 & 0.013 & 0.009 & 0.004 & 0.003 & 0.003\\
\textbf{t-iid (Vol)} & 0.032 & 0.026 & 0.018 & 0.012 & 0.010 & 0.007 & 0.003 & 0.003 & 0.002\\
\midrule
\textbf{t-MA (Clock)} & 0.036 & 0.035 & 0.033 & 0.013 & 0.013 & 0.013 & 0.003 & 0.003 & 0.004\\
\textbf{t-MA (TPV)} & 0.038 & 0.036 & 0.027 & 0.015 & 0.014 & 0.011 & 0.004 & 0.004 & 0.003\\
\textbf{t-MA (Vol)} & 0.033 & 0.029 & 0.019 & 0.012 & 0.011 & 0.008 & 0.003 & 0.003 & 0.002\\
\midrule
\textbf{t-iid (5; Clock)} & 0.055 & \underline{0.049} & \underline{0.046} & 0.024 & 0.022 & \underline{0.021} & \textbf{0.008} & 0.007 & \underline{0.006}\\
\textbf{t-iid (5; TPV)} & 0.055 & \textbf{0.050} & 0.034 & \textbf{0.025} & 0.023 & 0.016 & \underline{0.007} & 0.007 & 0.005\\
\textbf{t-iid (5; Vol)} & \underline{0.048} & 0.038 & 0.023 & 0.021 & 0.018 & 0.010 & 0.007 & 0.005 & 0.003\\
\midrule
\textbf{t-MA (5; Clock)} & 0.058 & 0.055 & \textbf{0.052} & \underline{0.025} & \underline{0.024} & \textbf{0.023} & 0.007 & \underline{0.007} & \textbf{0.007}\\
\textbf{t-MA (5; TPV)} & 0.059 & 0.056 & 0.039 & 0.026 & \textbf{0.026} & 0.017 & 0.007 & \textbf{0.008} & 0.006\\
\textbf{t-MA (5; Vol)} & \textbf{0.050} & 0.042 & 0.025 & 0.022 & 0.019 & 0.011 & 0.006 & 0.006 & 0.004\\
\midrule
\textbf{t-iid (21; Clock)} & 0.041 & 0.037 & 0.034 & 0.015 & 0.014 & 0.014 & 0.004 & 0.004 & 0.004\\
\textbf{t-iid (21; TPV)} & 0.042 & 0.038 & 0.027 & 0.016 & 0.016 & 0.011 & 0.005 & 0.004 & 0.003\\
\textbf{t-iid (21; Vol)} & 0.037 & 0.030 & 0.019 & 0.014 & 0.012 & 0.008 & 0.004 & 0.003 & 0.002\\
\midrule
\textbf{t-MA (21; Clock)} & 0.043 & 0.041 & 0.038 & 0.016 & 0.016 & 0.016 & 0.003 & 0.004 & 0.004\\
\textbf{t-MA (21; TPV)} & 0.045 & 0.042 & 0.030 & 0.018 & 0.017 & 0.012 & 0.004 & 0.005 & 0.004\\
\textbf{t-MA (21; Vol)} & 0.038 & 0.033 & 0.020 & 0.015 & 0.013 & 0.009 & 0.003 & 0.003 & 0.002\\
\bottomrule
\end{tabular}

\caption{Real-world dataset - hits frequency. Ideally, the hits frequency should be as close as possible to the target probability level $\theta$. The best result is in bold, the second to best is underlined. The subordinator is indicated inside the brackets.}
\label{tab:real_hits}
}
\end{center}
\end{table}

The statistical tests for ES are shown in Table \ref{tab:real_es}.
\begin{table}[H]
\begin{center}
\begin{adjustwidth}{-0.5cm}{}
{\small

\begin{tabular}{lcccccccccccc}
\toprule
\multicolumn{13}{c}{\textbf{BANK DATASET}} \\
\toprule
\multirow{4}{*}{\textbf{Algorithm}} & \multicolumn{6}{c}{$\theta=0.05$} & \multicolumn{6}{c}{$\theta=0.025$} \\
\cmidrule(lr){2-7} \cmidrule(lr){8-13}
 & \multicolumn{2}{c}{$c=39$} & \multicolumn{2}{c}{$c=78$} & \multicolumn{2}{c}{$c=130$} & \multicolumn{2}{c}{$c=39$} & \multicolumn{2}{c}{$c=78$} & \multicolumn{2}{c}{$c=130$} \\
\cmidrule(lr){2-3} \cmidrule(lr){4-5} \cmidrule(lr){6-7} \cmidrule(lr){8-9} \cmidrule(lr){10-11} \cmidrule(lr){12-13}
 & AS1 & AS2 & AS1 & AS2 & AS1 & AS2 & AS1 & AS2 & AS1 & AS2 & AS1 & AS2 \\
\midrule
\textbf{DH (Clock)} & 0.608 & 0.554 & 0.693 & 0.536 & 0.681 & 0.458 & 0.241 & 0.084 & 0.289 & 0.114 & 0.325 & 0.139\\
\textbf{DH (TPV)} & 0.331 & 0.380 & 0.380 & 0.410 & 0.380 & 0.566 & 0.120 & 0.060 & 0.211 & 0.078 & 0.253 & 0.078\\
\textbf{DH (Vol)} & 0.428 & 0.590 & 0.446 & 0.651 & 0.416 & 0.735 & 0.157 & 0.090 & 0.241 & 0.102 & 0.410 & 0.199\\
\midrule
\textbf{t-iid (Clock)} & 0.175 & 0.253 & 0.157 & 0.265 & 0.102 & 0.229 & 0.187 & 0.072 & 0.169 & \underline{0.054} & 0.193 & 0.060\\
\textbf{t-iid (TPV)} & 0.187 & 0.193 & 0.145 & 0.241 & \textbf{0.090} & 0.392 & 0.199 & 0.084 & 0.175 & \underline{0.054} & 0.217 & 0.072\\
\textbf{t-iid (Vol)} & 0.151 & 0.271 & \underline{0.120} & 0.367 & 0.151 & 0.500 & 0.211 & 0.072 & 0.253 & 0.060 & 0.380 & 0.175\\
\midrule
\textbf{t-MA (Clock)} & 0.289 & 0.223 & 0.247 & 0.205 & 0.187 & 0.229 & 0.223 & 0.066 & 0.205 & 0.060 & 0.205 & \textbf{0.054}\\
\textbf{t-MA (TPV)} & 0.229 & 0.175 & 0.193 & 0.187 & 0.157 & 0.367 & 0.187 & \textbf{0.036} & 0.163 & \textbf{0.048} & 0.163 & 0.090\\
\textbf{t-MA (Vol)} & 0.169 & 0.247 & 0.175 & 0.343 & 0.175 & 0.512 & 0.223 & 0.078 & 0.301 & 0.078 & 0.361 & 0.139\\
\midrule
\textbf{t-iid (5; Clock)} & 0.139 & 0.090 & \underline{0.120} & 0.120 & \textbf{0.090} & \textbf{0.108} & \textbf{0.096} & \textbf{0.036} & 0.133 & 0.066 & \textbf{0.120} & \textbf{0.054}\\
\textbf{t-iid (5; TPV)} & \underline{0.108} & \textbf{0.084} & \textbf{0.084} & \textbf{0.102} & \textbf{0.090} & 0.181 & \underline{0.108} & 0.060 & 0.151 & \underline{0.054} & 0.163 & 0.078\\
\textbf{t-iid (5; Vol)} & \textbf{0.090} & 0.090 & \underline{0.120} & 0.163 & 0.133 & 0.349 & 0.120 & 0.054 & \textbf{0.127} & 0.060 & 0.295 & 0.157\\
\midrule
\textbf{t-MA (5; Clock)} & 0.223 & 0.090 & 0.133 & 0.120 & 0.108 & \textbf{0.108} & 0.187 & 0.048 & 0.193 & 0.066 & \underline{0.127} & 0.072\\
\textbf{t-MA (5; TPV)} & 0.133 & \textbf{0.084} & 0.127 & \underline{0.114} & 0.096 & 0.175 & 0.157 & 0.048 & 0.157 & 0.060 & 0.133 & 0.060\\
\textbf{t-MA (5; Vol)} & 0.193 & 0.090 & \underline{0.120} & 0.157 & 0.145 & 0.361 & 0.139 & 0.042 & \textbf{0.127} & 0.072 & 0.325 & 0.139\\
\midrule
\textbf{t-iid (21; Clock)} & 0.211 & 0.157 & 0.193 & 0.199 & 0.139 & 0.211 & 0.187 & 0.072 & 0.181 & 0.072 & 0.181 & 0.072\\
\textbf{t-iid (21; TPV)} & 0.139 & 0.133 & 0.157 & 0.205 & 0.108 & 0.343 & 0.151 & 0.108 & 0.169 & 0.072 & 0.163 & 0.090\\
\textbf{t-iid (21; Vol)} & 0.181 & 0.187 & 0.133 & 0.307 & 0.193 & 0.440 & 0.181 & 0.042 & 0.163 & 0.084 & 0.380 & 0.193\\
\midrule
\textbf{t-MA (21; Clock)} & 0.289 & 0.139 & 0.247 & 0.163 & 0.181 & 0.217 & 0.151 & 0.066 & 0.193 & \underline{0.054} & 0.169 & 0.102\\
\textbf{t-MA (21; TPV)} & 0.265 & 0.090 & 0.175 & 0.127 & 0.127 & 0.295 & 0.175 & 0.072 & 0.163 & 0.084 & 0.157 & 0.090\\
\textbf{t-MA (21; Vol)} & 0.187 & 0.217 & 0.193 & 0.289 & 0.211 & 0.476 & 0.193 & 0.054 & 0.217 & 0.090 & 0.380 & 0.157\\
\midrule
\bottomrule
\end{tabular}

\vspace{0.1cm}

\begin{tabular}{lcccccccccccc}
\toprule
\multicolumn{13}{c}{\textbf{NASDAQ DATASET}} \\
\toprule
\multirow{4}{*}{\textbf{Algorithm}} & \multicolumn{6}{c}{$\theta=0.05$} & \multicolumn{6}{c}{$\theta=0.025$} \\
\cmidrule(lr){2-7} \cmidrule(lr){8-13}
 & \multicolumn{2}{c}{$c=39$} & \multicolumn{2}{c}{$c=78$} & \multicolumn{2}{c}{$c=130$} & \multicolumn{2}{c}{$c=39$} & \multicolumn{2}{c}{$c=78$} & \multicolumn{2}{c}{$c=130$} \\
\cmidrule(lr){2-3} \cmidrule(lr){4-5} \cmidrule(lr){6-7} \cmidrule(lr){8-9} \cmidrule(lr){10-11} \cmidrule(lr){12-13}
 & AS1 & AS2 & AS1 & AS2 & AS1 & AS2 & AS1 & AS2 & AS1 & AS2 & AS1 & AS2 \\
\midrule
\textbf{DH (Clock)} & 0.563 & 0.617 & 0.617 & 0.683 & 0.569 & 0.653 & 0.281 & 0.108 & 0.419 & 0.216 & 0.503 & 0.222\\
\textbf{DH (TPV)} & 0.186 & 0.575 & 0.222 & 0.599 & 0.275 & 0.665 & 0.144 & 0.060 & 0.317 & 0.168 & 0.413 & 0.180\\
\textbf{DH (Vol)} & 0.210 & 0.581 & 0.246 & 0.623 & 0.287 & 0.695 & 0.305 & 0.090 & 0.413 & 0.180 & 0.545 & 0.281\\
\midrule
\textbf{t-iid (Clock)} & 0.150 & 0.359 & 0.156 & 0.413 & 0.126 & 0.407 & 0.263 & 0.090 & 0.257 & 0.132 & 0.317 & 0.126\\
\textbf{t-iid (TPV)} & 0.174 & 0.323 & 0.138 & 0.377 & 0.138 & 0.521 & 0.210 & 0.102 & 0.204 & 0.114 & 0.359 & 0.156\\
\textbf{t-iid (Vol)} & 0.168 & 0.383 & 0.132 & 0.509 & 0.228 & 0.473 & 0.299 & 0.102 & 0.401 & 0.168 & 0.551 & 0.275\\
\midrule
\textbf{t-MA (Clock)} & 0.299 & 0.383 & 0.246 & 0.353 & 0.204 & 0.335 & 0.287 & 0.132 & 0.246 & 0.108 & 0.269 & 0.114\\
\textbf{t-MA (TPV)} & 0.180 & 0.275 & 0.162 & 0.341 & 0.126 & 0.521 & 0.216 & 0.054 & 0.186 & 0.114 & 0.293 & 0.132\\
\textbf{t-MA (Vol)} & 0.210 & 0.419 & 0.144 & 0.449 & 0.228 & 0.479 & 0.257 & 0.108 & 0.341 & 0.150 & 0.521 & 0.204\\
\midrule
\textbf{t-iid (5; Clock)} & 0.132 & \underline{0.048} & 0.084 & \textbf{0.054} & \textbf{0.090} & \underline{0.114} & \underline{0.096} & 0.030 & \textbf{0.090} & 0.036 & \textbf{0.102} & \underline{0.048}\\
\textbf{t-iid (5; TPV)} & \textbf{0.096} & 0.078 & \textbf{0.072} & 0.084 & \textbf{0.090} & 0.246 & 0.144 & \textbf{0.024} & \underline{0.126} & 0.042 & 0.192 & 0.084\\
\textbf{t-iid (5; Vol)} & \underline{0.102} & 0.096 & \textbf{0.072} & 0.263 & 0.174 & 0.335 & \textbf{0.078} & 0.042 & 0.216 & 0.042 & 0.425 & 0.180\\
\midrule
\textbf{t-MA (5; Clock)} & 0.186 & \textbf{0.036} & 0.150 & 0.072 & 0.144 & \textbf{0.060} & 0.162 & 0.036 & 0.144 & 0.030 & \underline{0.156} & \textbf{0.012}\\
\textbf{t-MA (5; TPV)} & 0.180 & 0.072 & 0.120 & \underline{0.060} & 0.096 & 0.240 & 0.192 & \textbf{0.024} & 0.132 & \textbf{0.024} & 0.162 & 0.078\\
\textbf{t-MA (5; Vol)} & 0.216 & 0.090 & 0.102 & 0.275 & 0.186 & 0.377 & 0.168 & 0.048 & 0.210 & \textbf{0.024} & 0.389 & 0.192\\
\midrule
\textbf{t-iid (21; Clock)} & 0.204 & 0.222 & 0.186 & 0.287 & 0.120 & 0.323 & 0.180 & 0.078 & 0.204 & 0.084 & 0.198 & 0.144\\
\textbf{t-iid (21; TPV)} & 0.180 & 0.174 & 0.180 & 0.216 & 0.114 & 0.479 & 0.210 & 0.084 & 0.168 & 0.042 & 0.287 & 0.138\\
\textbf{t-iid (21; Vol)} & 0.192 & 0.323 & 0.174 & 0.401 & 0.251 & 0.461 & 0.162 & 0.102 & 0.251 & 0.156 & 0.473 & 0.228\\
\midrule
\textbf{t-MA (21; Clock)} & 0.353 & 0.168 & 0.234 & 0.228 & 0.210 & 0.251 & 0.174 & 0.108 & 0.269 & 0.066 & 0.174 & 0.066\\
\textbf{t-MA (21; TPV)} & 0.251 & 0.138 & 0.186 & 0.198 & 0.144 & 0.449 & 0.168 & 0.060 & 0.150 & 0.054 & 0.216 & 0.108\\
\textbf{t-MA (21; Vol)} & 0.222 & 0.287 & 0.174 & 0.359 & 0.216 & 0.425 & 0.240 & 0.078 & 0.275 & 0.096 & 0.449 & 0.210\\
\bottomrule
\end{tabular}
}

\end{adjustwidth}
\caption{Real-world dataset - rejection frequency for the AS1 and AS2 tests. The null is that the estimated pair $(\hat{q}, \hat{e})$ is a good proxy for (VaR, ES). The best result is in bold, the second to best is underlined. The subordinator is indicated inside the brackets.}
\label{tab:real_es}
\end{center}
\end{table}
All our filtering strategies outperform \textbf{DH} in most of the tests. The comparison between subordinators provides similar insights to those obtained when discussing quantile results. Indeed, neither Clock nor TPV strongly outperforms the other, and Vol exhibits poor performance. It is worth discussing the behavior of the rejection rate as a function of the probability level $\theta$. As previously observed, the results on the synthetic dataset exhibit a decay as $\theta$ becomes smaller. Instead, in this context, we observe a decrease in the rejection rate for the AS2 test. This is probably due to the fact that we use in-sample observations. Finally, no specific patterns can be extracted about the rate behavior as a function of $c$.\\

As for the out-of-sample results, they have been estimated as the forecasting loss achieved by an AR(1) model built on top of the estimated risk measures. The results for the VaR, measured in terms of the Pinball loss described in Eq. \eqref{eq:1113_0900}, are in Table \ref{tab:real_loss_q}. Again, the proposed filters usually show better performances than \textbf{DH}, and the $EMA(21)$ version usually displays the best results. As for the subordinators, there are no substantial differences from the in-sample results. Instead, there is a discrepancy regarding the MA filter, as the best performances are generally obtained without using it. Moreover, the loss seems to be unaffected by the choice of $c$. That is, sometimes $c=39$ is the best choice, and other times $c=130$ performs better, and there are no noticeable patterns.\\
\begin{table}[ht!]
\begin{center}
{\small

\begin{tabular}{lccccccccc}
\toprule
\multicolumn{10}{c}{\textbf{BANK DATASET}} \\
\toprule
\multirow{2}{*}{\textbf{Algorithm}} & \multicolumn{3}{c}{$\theta=0.05$} & \multicolumn{3}{c}{$\theta=0.025$} & \multicolumn{3}{c}{$\theta=0.01$} \\
\cmidrule(lr){2-4} \cmidrule(lr){5-7} \cmidrule(lr){8-10}
 & $c=39$ & $c=78$ & $c=130$ & $c=39$ & $c=78$ & $c=130$ & $c=39$ & $c=78$ & $c=130$ \\
\midrule
\textbf{DH (Clock)} & 2.016 & 1.927 & 1.885 & 1.284 & 1.199 & 1.147 & 0.758 & 0.65 & 0.598\\
\textbf{DH (TPV)} & 1.962 & 1.902 & \underline{1.859} & 1.254 & 1.173 & 1.14 & 0.772 & 0.721 & 0.675\\
\textbf{DH (Vol)} & 1.976 & 1.899 & 1.866 & 1.251 & 1.17 & 1.145 & 0.754 & 0.671 & 0.636\\
\midrule
\textbf{t-iid (Clock)} & 1.887 & 1.876 & 1.873 & 1.142 & 1.131 & 1.124 & 0.587 & 0.579 & 0.563\\
\textbf{t-iid (TPV)} & 1.874 & 1.885 & 1.917 & 1.138 & 1.142 & 1.153 & 0.585 & 0.59 & 0.579\\
\textbf{t-iid (Vol)} & 1.908 & 1.927 & 1.951 & 1.169 & 1.175 & 1.184 & 0.612 & 0.615 & 0.607\\
\midrule
\textbf{t-MA (Clock)} & 1.877 & 1.869 & 1.874 & 1.129 & 1.124 & \underline{1.116} & 0.57 & 0.564 & 0.554\\
\textbf{t-MA (TPV)} & 1.864 & 1.891 & 1.934 & 1.12 & 1.151 & 1.163 & 0.568 & 0.593 & 0.583\\
\textbf{t-MA (Vol)} & 1.888 & 1.911 & 1.952 & 1.14 & 1.151 & 1.17 & 0.581 & 0.586 & 0.589\\
\midrule
\textbf{t-iid (5; Clock)} & 1.938 & 1.93 & 1.929 & 1.166 & 1.154 & 1.151 & 0.583 & 0.582 & 0.576\\
\textbf{t-iid (5; TPV)} & 1.924 & 1.931 & 1.962 & 1.158 & 1.16 & 1.172 & 0.592 & 0.593 & 0.584\\
\textbf{t-iid (5; Vol)} & 1.945 & 1.964 & 1.993 & 1.176 & 1.19 & 1.198 & 0.609 & 0.614 & 0.606\\
\midrule
\textbf{t-MA (5; Clock)} & 1.902 & 1.908 & 1.902 & 1.132 & 1.139 & 1.132 & 0.563 & 0.565 & \textbf{0.55}\\
\textbf{t-MA (5; TPV)} & 1.9 & 1.925 & 1.972 & 1.132 & 1.16 & 1.172 & 0.568 & 0.588 & 0.588\\
\textbf{t-MA (5; Vol)} & 1.915 & 1.935 & 1.971 & 1.145 & 1.16 & 1.171 & 0.574 & 0.581 & 0.575\\
\midrule
\textbf{t-iid (21; Clock)} & 1.867 & \underline{1.86} & \textbf{1.855} & 1.128 & \underline{1.114} & \textbf{1.106} & 0.569 & \underline{0.564} & 0.552\\
\textbf{t-iid (21; TPV)} & \underline{1.855} & 1.871 & 1.908 & 1.114 & 1.123 & 1.138 & 0.57 & 0.577 & 0.567\\
\textbf{t-iid (21; Vol)} & 1.889 & 1.909 & 1.936 & 1.142 & 1.153 & 1.166 & 0.595 & 0.599 & 0.593\\
\midrule
\textbf{t-MA (21; Clock)} & 1.859 & \textbf{1.859} & 1.869 & \underline{1.111} & \textbf{1.109} & 1.119 & \underline{0.559} & \textbf{0.557} & \underline{0.552}\\
\textbf{t-MA (21; TPV)} & \textbf{1.848} & 1.882 & 1.927 & \textbf{1.103} & 1.13 & 1.15 & \textbf{0.556} & 0.575 & 0.575\\
\textbf{t-MA (21; Vol)} & 1.867 & 1.897 & 1.943 & 1.12 & 1.14 & 1.163 & 0.565 & 0.575 & 0.577\\
\bottomrule
\end{tabular}

\vspace{0.1cm}

\begin{tabular}{lccccccccc}
\toprule
\multicolumn{10}{c}{\textbf{NASDAQ DATASET}} \\
\toprule
\multirow{2}{*}{\textbf{Algorithm}} & \multicolumn{3}{c}{$\theta=0.05$} & \multicolumn{3}{c}{$\theta=0.025$} & \multicolumn{3}{c}{$\theta=0.01$} \\
\cmidrule(lr){2-4} \cmidrule(lr){5-7} \cmidrule(lr){8-10}
 & $c=39$ & $c=78$ & $c=130$ & $c=39$ & $c=78$ & $c=130$ & $c=39$ & $c=78$ & $c=130$ \\
\midrule
\textbf{DH (Clock)} & 1.893 & 1.846 & 1.822 & 1.167 & 1.129 & 1.092 & 0.613 & 0.569 & 0.548\\
\textbf{DH (TPV)} & 1.853 & 1.817 & \textbf{1.801} & 1.143 & 1.101 & 1.081 & 0.621 & 0.566 & 0.55\\
\textbf{DH (Vol)} & 1.853 & \textbf{1.815} & \underline{1.803} & 1.132 & 1.091 & \underline{1.081} & 0.601 & 0.557 & 0.548\\
\midrule
\textbf{t-iid (Clock)} & 1.83 & 1.834 & 1.823 & 1.094 & 1.095 & 1.082 & 0.549 & 0.543 & 0.536\\
\textbf{t-iid (TPV)} & \underline{1.812} & 1.82 & 1.88 & \textbf{1.074} & \underline{1.079} & 1.118 & \textbf{0.529} & \textbf{0.531} & 0.552\\
\textbf{t-iid (Vol)} & 1.833 & 1.856 & 1.919 & 1.092 & 1.103 & 1.145 & 0.542 & 0.544 & 0.566\\
\midrule
\textbf{t-MA (Clock)} & 1.829 & 1.827 & 1.827 & 1.093 & 1.086 & 1.085 & 0.545 & 0.535 & \textbf{0.532}\\
\textbf{t-MA (TPV)} & 1.827 & 1.832 & 1.901 & 1.082 & 1.084 & 1.127 & 0.537 & 0.539 & 0.555\\
\textbf{t-MA (Vol)} & 1.835 & 1.868 & 1.942 & 1.089 & 1.111 & 1.162 & 0.539 & 0.546 & 0.573\\
\midrule
\textbf{t-iid (5; Clock)} & 1.855 & 1.859 & 1.854 & 1.106 & 1.109 & 1.102 & 0.548 & 0.544 & 0.542\\
\textbf{t-iid (5; TPV)} & 1.843 & 1.854 & 1.892 & 1.098 & 1.1 & 1.122 & 0.54 & 0.539 & 0.55\\
\textbf{t-iid (5; Vol)} & 1.856 & 1.874 & 1.925 & 1.106 & 1.113 & 1.144 & 0.547 & 0.546 & 0.56\\
\midrule
\textbf{t-MA (5; Clock)} & 1.856 & 1.858 & 1.87 & 1.105 & 1.101 & 1.109 & 0.547 & 0.536 & 0.541\\
\textbf{t-MA (5; TPV)} & 1.854 & 1.87 & 1.917 & 1.099 & 1.106 & 1.134 & 0.541 & 0.54 & 0.554\\
\textbf{t-MA (5; Vol)} & 1.86 & 1.884 & 1.958 & 1.101 & 1.119 & 1.168 & 0.539 & 0.549 & 0.572\\
\midrule
\textbf{t-iid (21; Clock)} & 1.822 & 1.825 & 1.813 & 1.091 & 1.09 & \textbf{1.08} & 0.543 & 0.539 & 0.535\\
\textbf{t-iid (21; TPV)} & \textbf{1.808} & \underline{1.816} & 1.865 & \underline{1.075} & \textbf{1.078} & 1.109 & \underline{0.53} & \underline{0.531} & 0.546\\
\textbf{t-iid (21; Vol)} & 1.827 & 1.845 & 1.902 & 1.089 & 1.098 & 1.134 & 0.54 & 0.541 & 0.559\\
\midrule
\textbf{t-MA (21; Clock)} & 1.825 & 1.827 & 1.828 & 1.091 & 1.088 & 1.086 & 0.542 & 0.535 & \underline{0.534}\\
\textbf{t-MA (21; TPV)} & 1.826 & 1.834 & 1.891 & 1.084 & 1.087 & 1.121 & 0.537 & 0.538 & 0.552\\
\textbf{t-MA (21; Vol)} & 1.836 & 1.862 & 1.93 & 1.092 & 1.107 & 1.155 & 0.538 & 0.544 & 0.569\\
\bottomrule
\end{tabular}

\caption{Real-world dataset - VaR loss function for regression. The estimated $\hat{q}$ is used to train an AR(1) model, whose forecast is used as the VaR estimate for the next day. The subordinator is indicated inside the brackets.}
\label{tab:real_loss_q}
}
\end{center}
\end{table}

Finally, the results for the ES, computed using the loss in Eq. \eqref{eq:1113_0901}, are in Table \ref{tab:real_loss_e}. Few comments are in order.  First, the best results are obtained by our approach with the $EMA(21)$ drift. As for the Bank dataset, the iid specification outperforms the MA-based one, while the opposite is true in the Nasdaq dataset. Then, we find that the Clock and TPV subordinators outperform the others, and there is no clear indication of what is best. The impact of $c$ is still not unidirectional. Moreover, we recover the increase in loss as $\theta$ becomes smaller, as found in \cite{gatta2024caesar}. This confirms our hypothesis that the degradation is due to the high unpredictability of the events corresponding to the extreme values in the deep tail. Finally, further experiments have been carried out with different forecasters, namely the Exponential Moving Average and the Random Walk. The results are detailed in Appendix \ref{app:for_rrm}. It is worth highlighting that they are very close to what is shown in Table \ref{tab:real_loss_q} and \ref{tab:real_loss_e}, thus confirming our findings.
\begin{table}[H]
\begin{center}
\begin{adjustwidth}{-1.2cm}{}
{\small

\begin{tabular}{lcccccc||cccccc}
\toprule
 & \multicolumn{6}{c}{BANK DATASET} & \multicolumn{6}{c}{NASDAQ DATASET} \\
\cmidrule(lr){2-7} \cmidrule(lr){8-13}
\multirow{2}{*}{\textbf{Algorithm}} & \multicolumn{3}{c}{$\theta=0.05$} & \multicolumn{3}{c}{$\theta=0.025$} & \multicolumn{3}{c}{$\theta=0.05$} & \multicolumn{3}{c}{$\theta=0.025$} \\
\cmidrule(lr){2-4} \cmidrule(lr){5-7} \cmidrule(lr){8-10} \cmidrule(lr){11-13}
 & $c=39$ & $c=78$ & $c=130$ & $c=39$ & $c=78$ & $c=130$ & $c=39$ & $c=78$ & $c=130$ & $c=39$ & $c=78$ & $c=130$ \\
\midrule
\textbf{DH (Clock)} & 1.284 & 1.214 & 1.182 & 1.566 & 1.451 & 1.391 & 1.247 & 1.209 & \underline{1.189} & 1.476 & 1.42 & 1.381\\
\textbf{DH (TPV)} & 1.258 & 1.217 & 1.197 & 1.556 & 1.467 & 1.425 & 1.224 & 1.196 & 1.204 & 1.463 & 1.403 & 1.401\\
\textbf{DH (Vol)} & 1.262 & 1.218 & 1.206 & 1.546 & 1.459 & 1.43 & 1.222 & 1.201 & 1.215 & 1.44 & 1.401 & 1.415\\
\midrule
\textbf{t-iid (Clock)} & 1.19 & 1.183 & 1.177 & 1.386 & 1.376 & 1.364 & 1.201 & 1.2 & 1.192 & 1.387 & 1.383 & 1.371\\
\textbf{t-iid (TPV)} & 1.173 & 1.179 & 1.206 & 1.368 & 1.371 & 1.392 & \textbf{1.183} & \underline{1.189} & 1.229 & \textbf{1.361} & \textbf{1.365} & 1.406\\
\textbf{t-iid (Vol)} & 1.201 & 1.215 & 1.238 & 1.406 & 1.416 & 1.433 & 1.202 & 1.22 & 1.265 & 1.386 & 1.399 & 1.448\\
\midrule
\textbf{t-MA (Clock)} & 1.18 & 1.176 & 1.172 & 1.368 & 1.364 & 1.351 & 1.196 & 1.196 & 1.192 & 1.379 & 1.376 & \textbf{1.368}\\
\textbf{t-MA (TPV)} & 1.166 & 1.184 & 1.205 & 1.353 & 1.378 & 1.386 & 1.193 & 1.198 & 1.238 & 1.374 & 1.376 & 1.412\\
\textbf{t-MA (Vol)} & 1.189 & 1.2 & 1.232 & 1.38 & 1.387 & 1.417 & 1.203 & 1.225 & 1.275 & 1.384 & 1.405 & 1.459\\
\midrule
\textbf{t-iid (5; Clock)} & 1.222 & 1.214 & 1.215 & 1.407 & 1.396 & 1.395 & 1.221 & 1.219 & 1.215 & 1.402 & 1.402 & 1.395\\
\textbf{t-iid (5; TPV)} & 1.21 & 1.207 & 1.227 & 1.398 & 1.393 & 1.407 & 1.217 & 1.214 & 1.234 & 1.401 & 1.392 & 1.409\\
\textbf{t-iid (5; Vol)} & 1.224 & 1.236 & 1.255 & 1.42 & 1.431 & 1.443 & 1.219 & 1.229 & 1.26 & 1.402 & 1.407 & 1.44\\
\midrule
\textbf{t-MA (5; Clock)} & 1.197 & 1.199 & 1.194 & 1.375 & 1.377 & 1.368 & 1.218 & 1.214 & 1.22 & 1.397 & 1.39 & 1.399\\
\textbf{t-MA (5; TPV)} & 1.192 & 1.201 & 1.219 & 1.373 & 1.388 & 1.394 & 1.217 & 1.224 & 1.245 & 1.4 & 1.401 & 1.418\\
\textbf{t-MA (5; Vol)} & 1.202 & 1.213 & 1.237 & 1.386 & 1.396 & 1.415 & 1.218 & 1.229 & 1.274 & 1.394 & 1.409 & 1.457\\
\midrule
\textbf{t-iid (21; Clock)} & 1.178 & 1.17 & \underline{1.166} & 1.373 & 1.357 & \underline{1.349} & 1.197 & 1.195 & \textbf{1.186} & 1.384 & 1.379 & \underline{1.369}\\
\textbf{t-iid (21; TPV)} & \underline{1.162} & \underline{1.168} & 1.194 & \underline{1.35} & \underline{1.355} & 1.378 & \underline{1.185} & \textbf{1.187} & 1.218 & \underline{1.367} & \underline{1.366} & 1.395\\
\textbf{t-iid (21; Vol)} & 1.185 & 1.198 & 1.224 & 1.384 & 1.393 & 1.417 & 1.198 & 1.211 & 1.251 & 1.382 & 1.393 & 1.434\\
\midrule
\textbf{t-MA (21; Clock)} & 1.17 & \textbf{1.167} & \textbf{1.166} & 1.355 & \textbf{1.35} & \textbf{1.346} & 1.195 & 1.195 & 1.191 & 1.379 & 1.377 & 1.37\\
\textbf{t-MA (21; TPV)} & \textbf{1.157} & 1.173 & 1.197 & \textbf{1.338} & 1.359 & 1.376 & 1.194 & 1.2 & 1.229 & 1.378 & 1.381 & 1.403\\
\textbf{t-MA (21; Vol)} & 1.175 & 1.188 & 1.222 & 1.363 & 1.373 & 1.406 & 1.203 & 1.219 & 1.263 & 1.385 & 1.399 & 1.449\\
\bottomrule
\end{tabular}

}
\end{adjustwidth}
\caption{Real-world dataset - ES loss function for regression. The estimated pair $(\hat{q}, \hat{e})$ is used to train two AR(1) models, whose forecasts are used as an estimate for the pair (VaR, ES) for the next day.  The subordinator is indicated inside the brackets.}
\label{tab:real_loss_e}
\end{center}
\end{table}
\section{Conclusion}
\label{sec:conc}
In this paper, we deal with the problem of estimating risk measures, which can be considered as latent quantities. Our approach relies on using high-frequency (minute-by-minute) information to estimate the low-frequency (daily) risk. In this sense, our proposal is similar to the risk forecasting based on Realized Volatility, so we named it {\it Realized} Risk Measures. The key difference concerning RV is the parametric assumption needed to model the tail of the return distribution. Despite the underlying idea being quite simple, there is only one similar work proposed in \cite{dimitriadis2022realized}, which is based on the assumption of self-similarity of subordinated prices. However this assumption is often not fully supported by empirical data and can lead to suboptimal estimations of risk. We aim to overcome the limitation of \cite{dimitriadis2022realized} by better tailoring the proposed method to empirical financial returns. Namely, we take into account stylized facts as fat-tails and microstructural effects in terms of autocorrelation in data. Furthermore, we also discuss an estimation bias related to non-zero empirical mean of the observed high-frequency returns.\\

For validation, we have compared the proposed method with the benchmark proposed by \cite{dimitriadis2022realized} using both synthetic and real-world datasets. In the first case, the comparison has been done by looking at the distance between the estimated risk measure and the ground truth. In the second case, we have studied both the in-sample performance using statistical tests and the out-of-sample performance by building an autoregressive model on top of the realized risk and evaluating the forecasts with popular loss functions. Our approach outperforms the benchmark in all cases.\\

As for future research, we see two potential directions. First, we aim to study the effect of including jumps in the high-frequency dynamics. Theoretically, one could expect an improvement in the estimated low-frequency measures. However, empirical works in the literature raise some doubts -Ref. \cite{vzikevs2015semi} argues that the jump measure coefficient is not statistically significant. Similarly,  Ref. \cite{clements2008quantile} finds that the HAR model with or without a separate jump component does not lead to significant improvements. For this reason, we have not included jumps in the current work, and we want to deepen this analysis in the future. Last, we aim to use the RRM as an additional feature for risk forecasting. Indeed, we expect this could enhance state-of-the-art approaches as the RRM contains high-frequency information that traditional forecasters neglect.


%
%
\section*{Data and Code Availability}
The data utilized in this paper are not available due to copyright restrictions. The code is publicly accessible at the GitHub repository \url{https://github.com/fgt996/Realized_Risk_Measures}.
\section*{Acknowledgments}
We would like to thank the organizers and participants of the Quantitative Finance Workshop 2025 and of the International Fintech Research Conference. PM acknowledges Gianna Fig\'a-Talamanca and Marco Patacca for useful discussions. PM acknowledges financial support under the National Recovery and Resilience Plan (PNNR), Mission 4, Component 2, Investment 1.1, Call for tender No. 104 pub-
lished on 2.2.2022 by the Italian Ministry of University and Research (MUR), funded by the European Union – NextGenerationEU– Project Title ‘‘Realized Random Graphs: A New Econometric Methodology
for the Inference of Dynamic Networks’’ – CUP B53D23010100001 - Grant Assignment Decree No. 2022MRSYB7 by the Italian Ministry of University and Research (MUR).

\bibliographystyle{abbrv}
\bibliography{Bibliography}
\appendix
\section{Implementation Details}
\label{app_imp_det}
In this appendix, we briefly discuss some details about the implementation of the proposed methodology. First, as for fitting the intra-day Student's t-distribution, we cap the degrees of freedom $\nu$ and the standard deviation $\sigma$. $\nu$ is forced to be greater or equal to $2+10^{-6}$, as the variance of the t distribution is defined only when $\nu>2$. $\sigma$ is forced to be greater than or equal to $10^{-6}$.\\

As for the Monte-Carlo approach for aggregation, the antithetic variates method for variance reduction has been used. Regarding the characteristic function approach, we have used the \texttt{QUADPACK} routine \cite{piessens2012quadpack} for numerical quadrature. Instead, the zero-finding function is mainly based on Brent's method. The target function $f$ is the difference between the CDF given by the Gil-Pelaez theorem (as in Eq. \eqref{eq:1217_1340}) and the probability level $\theta$. The key points are the following:
\begin{enumerate}
    \item Set a starting point $x_0=-10^{-3}$, desired tolerance $tol=10^{-8}$ and maximum number of iterations $n=10$.
    \item Find a suitable range $[a,b]$ such that $f(a)f(b) < 0$. $a$ are $b$ are searched by trials with a loop on ten log-equispaced points between: $-0.2$ and $x_0$, for $a$ candidates; $x_0$ and $10^{-12}$ for $b$ candidates. The priority is given to the closer pair $(a,b)$.
    \item Run the Brent's method. If the desired tolerance is achieved, its result is returned as the output. Otherwise, reiterate using the previous iteration's result as one of the boundaries. Repeat until convergence, or the maximum number of iterations is reached.
    \item If the tolerance has not been achieved yet, we use the Sequential Least Squares algorithm. The target function is $|f|$, the starting point is $x_0$, the domain is $[0,1]$. Different tolerance levels on the proposed solution are tried until we have $|f|<10^{-4}$.
\end{enumerate}
An overview of the computational time required for estimating the risk from one day of data is provided in Table \ref{tab:comp_time}. The table shows the mean $\pm$ one standard deviation computed over 252 iterations. The code runs on AMD Ryzen Threadripper PRO 5975WX.
\begin{table}[h]
\begin{center}
{\small
\begin{tabular}{ccccccc}
\toprule
\textbf{TPV Sub.} & \textbf{Fit t (iid)} & \textbf{Ch. Fun. (iid)} & \textbf{MC (iid)} & \textbf{Fit t(MA)} & \textbf{Ch. Fun. (MA)} & \textbf{MC (MA)}\\
\midrule
45.4 $\pm$ 1.5 & 13.7 $\pm$ 6.0 & 623.5 $\pm$ 47.5 & 195.5 $\pm$ 3.1 & 269.3 $\pm$ 106.1 & 626.9 $\pm$ 49.3 & 198.4 $\pm$ 3.2 \\
\bottomrule
\end{tabular}
\caption{Computational times, in milliseconds. Those of Clock and Vol subordinators are approximated to zero (in the milliseconds time scale), so they have been neglected. \textbf{Ch. Fun.} and \textbf{MC} are for the characteristic function and Monte-Carlo approach. \textbf{iid} and \textbf{MA} refer to the absence or presence of the MA filter.}
\label{tab:comp_time}
}
\end{center}
\end{table}
\section{Further Insights from the Simulation Study}
\label{app:loss_bias}
As stated in Section \ref{subsec:ev_comp} of the main text, the commonly used loss functions for forecasting problems lose their effectiveness when used in-sample. This can be empirically confirmed by comparing the loss of the estimated risk measures and the ground truth risk (see Section \ref{sec:exp} of the main text for the synthetic dataset description) in Table \ref{tab:sim_q_loss}, which shows the Pinball loss $\mathcal{L}_q^\theta$ for quantiles. The results in the table indicate that the ground truth is in the worst-performing approaches. A similar conclusion can be drawn by looking at Table \ref{tab:sim_e_loss}, where the $\mathcal{L}_e^\theta$ for the pair (VaR, ES) is shown.\\
\begin{table}[h]
\begin{center}
{\small

\begin{tabular}{lccccccccc}
\toprule
\multicolumn{10}{c}{\textbf{GAUSSIAN SYNTHETIC DATASET}} \\
\toprule
\multirow{2}{*}{\textbf{Algorithm}} & \multicolumn{3}{c}{$\theta=0.05$} & \multicolumn{3}{c}{$\theta=0.025$} & \multicolumn{3}{c}{$\theta=0.01$} \\
\cmidrule(lr){2-4} \cmidrule(lr){5-7} \cmidrule(lr){8-10}
 & $c=39$ & $c=78$ & $c=130$ & $c=39$ & $c=78$ & $c=130$ & $c=39$ & $c=78$ & $c=130$ \\
\midrule
\textbf{DH} & \textbf{1.146} & \textbf{1.242} & \textbf{1.267} & \textbf{0.659} & \textbf{0.709} & \textbf{0.711} & \textbf{0.314} & \textbf{0.328} & \textbf{0.326}\\
\textbf{t-iid} & 1.269 & 1.333 & 1.351 & 0.721 & 0.758 & 0.760 & 0.332 & 0.344 & 0.340\\
\textbf{t-MA} & \underline{1.238} & \underline{1.317} & \underline{1.337} & \underline{0.699} & \underline{0.745} & \underline{0.750} & \underline{0.320} & \underline{0.334} & \underline{0.338}\\
\textbf{t-iid (5)} & 1.372 & 1.463 & 1.503 & 0.767 & 0.834 & 0.860 & 0.347 & 0.381 & 0.388\\
\textbf{t-MA (5)} & 1.341 & 1.442 & 1.490 & 0.742 & 0.811 & 0.849 & 0.333 & 0.366 & 0.381\\
\textbf{t-iid (21)} & 1.290 & 1.364 & 1.390 & 0.730 & 0.779 & 0.784 & 0.332 & 0.352 & 0.351\\
\textbf{t-MA (21)} & 1.259 & 1.349 & 1.377 & 0.704 & 0.761 & 0.770 & 0.320 & 0.345 & 0.347\\
\textbf{GT} & 1.276 & 1.339 & 1.346 & 0.727 & 0.769 & 0.759 & 0.337 & 0.354 & 0.343\\
\bottomrule
\end{tabular}

\vspace{0.1cm}

\begin{tabular}{lccccccccc}
\toprule
\multicolumn{10}{c}{\textbf{STUDENT'S T SYNTHETIC DATASET}} \\
\toprule
\multirow{2}{*}{\textbf{Algorithm}} & \multicolumn{3}{c}{$\theta=0.05$} & \multicolumn{3}{c}{$\theta=0.025$} & \multicolumn{3}{c}{$\theta=0.01$} \\
\cmidrule(lr){2-4} \cmidrule(lr){5-7} \cmidrule(lr){8-10}
 & $c=39$ & $c=78$ & $c=130$ & $c=39$ & $c=78$ & $c=130$ & $c=39$ & $c=78$ & $c=130$ \\
\midrule
\textbf{DH} & \textbf{2.210} & \underline{2.883} & 3.085 & \textbf{1.164} & \textbf{1.685} & \textbf{1.766} & \textbf{0.476} & \textbf{0.535} & \textbf{0.903}\\
\textbf{t-iid} & \underline{2.461} & \textbf{2.865} & \textbf{2.892} & \underline{1.563} & \underline{1.835} & \underline{1.793} & \underline{0.912} & \underline{1.046} & \underline{0.961}\\
\textbf{t-MA} & 2.493 & 2.904 & \underline{2.911} & 1.599 & 1.871 & 1.823 & 0.936 & 1.072 & 0.982\\
\textbf{t-iid (5)} & 2.765 & 3.222 & 3.283 & 1.746 & 2.036 & 2.058 & 0.991 & 1.170 & 1.119\\
\textbf{t-MA (5)} & 2.785 & 3.248 & 3.299 & 1.750 & 2.057 & 2.080 & 1.001 & 1.182 & 1.120\\
\textbf{t-iid (21)} & 2.512 & 2.909 & 2.936 & 1.588 & 1.843 & 1.816 & 0.924 & 1.052 & 0.970\\
\textbf{t-MA (21)} & 2.540 & 2.943 & 2.963 & 1.613 & 1.880 & 1.848 & 0.946 & 1.072 & 0.987\\
\textbf{GT} & 2.740 & 3.044 & 3.034 & 1.848 & 2.012 & 1.940 & 1.114 & 1.199 & 1.092\\
\bottomrule
\end{tabular}

\caption{Synthetic dataset - comparison of Pinball losses ($\times 10^3$). The best result is bold, and the second to best is underlined. \textbf{t-iid} and \textbf{t-MA} are intended to be the approach based on t-distribution that averages the output from the characteristic function approach and the Monte-Carlo simulation for scaling the data to the daily frequency. \textbf{(GT)} stands for ground truth quantile.}
\label{tab:sim_q_loss}
}
\end{center}
\end{table}\\

\begin{table}[h]
\begin{center}
\begin{adjustwidth}{-0.9cm}{}
{\small

\begin{tabular}{lcccccc||cccccc}
\toprule
 & \multicolumn{6}{c}{\textbf{GAUSSIAN SYNTHETIC DATASET}} & \multicolumn{6}{c}{\textbf{STUDENT'S T SYNTHETIC DATASET}} \\
\cmidrule(lr){2-7} \cmidrule(lr){8-13}
\multirow{2}{*}{\textbf{Algorithm}} & \multicolumn{3}{c}{$\theta=0.05$} & \multicolumn{3}{c}{$\theta=0.025$} & \multicolumn{3}{c}{$\theta=0.05$} & \multicolumn{3}{c}{$\theta=0.025$} \\
\cmidrule(lr){2-4} \cmidrule(lr){5-7} \cmidrule(lr){8-10} \cmidrule(lr){11-13}
 & $c=39$ & $c=78$ & $c=130$ & $c=39$ & $c=78$ & $c=130$ & $c=39$ & $c=78$ & $c=130$ & $c=39$ & $c=78$ & $c=130$ \\
\midrule
\textbf{DH} & \textbf{0.815} & \textbf{0.904} & \textbf{0.928} & \textbf{0.957} & \textbf{1.039} & \textbf{1.046} & \textbf{1.065} & \textbf{1.289} & \textbf{1.414} & \textbf{1.233} & \textbf{1.435} & \textbf{1.522}\\
\textbf{t-iid} & 0.930 & 0.983 & 1.001 & 1.056 & 1.110 & 1.118 & \underline{1.482} & \underline{1.683} & \underline{1.719} & \underline{1.671} & \underline{1.888} & \underline{1.905}\\
\textbf{t-MA} & \underline{0.897} & \underline{0.966} & \underline{0.988} & \underline{1.018} & \underline{1.089} & \underline{1.102} & 1.493 & 1.701 & 1.722 & 1.692 & 1.911 & 1.917\\
\textbf{t-iid (5)} & 1.052 & 1.140 & 1.158 & 1.146 & 1.254 & 1.284 & 1.925 & 2.724 & 2.100 & 1.907 & 2.253 & 2.139\\
\textbf{t-MA (5)} & 1.013 & 1.106 & 1.149 & 1.097 & 1.206 & 1.267 & 1.771 & 2.076 & 2.027 & 1.837 & 2.091 & 2.288\\
\textbf{t-iid (21)} & 0.947 & 1.013 & 1.036 & 1.067 & 1.142 & 1.152 & 1.552 & 1.748 & 2.413 & 1.722 & 1.929 & 1.973\\
\textbf{t-MA (21)} & 0.913 & 0.996 & 1.022 & 1.023 & 1.113 & 1.131 & 1.543 & 1.749 & 1.927 & 1.723 & 1.961 & 1.978\\
\textbf{GT} & 0.939 & 0.992 & 0.998 & 1.069 & 1.129 & 1.118 & 1.712 & 1.832 & 1.827 & 2.006 & 2.106 & 2.071\\
\bottomrule
\end{tabular}

\caption{Synthetic dataset - (VaR, ES) joint loss.}
\label{tab:sim_e_loss}
}
\end{adjustwidth}
\end{center}
\end{table}

Next, we show the in-sample tests for the synthetic datasets. Directly studying the distance from the loss is preferable, but looking at the in-sample tests could be useful for comparing these results with real-world ones. Table \ref{tab:sim_q_intests} refers to the number of hits; Table \ref{tab:sim_e_intests} shows the AS tests for ES. Interestingly, \textbf{t-iid} and \textbf{t-MA} are the top-performing approaches, roughly confirming the intuition from the main text.
\begin{table}[h]
\begin{center}

{\small

\begin{tabular}{lccccccccc}
\toprule
\multicolumn{10}{c}{\textbf{GAUSSIAN SYNTHETIC DATASET}} \\
\toprule
\multirow{2}{*}{\textbf{Algorithm}} & \multicolumn{3}{c}{$\theta=0.05$} & \multicolumn{3}{c}{$\theta=0.025$} & \multicolumn{3}{c}{$\theta=0.01$} \\
\cmidrule(lr){2-4} \cmidrule(lr){5-7} \cmidrule(lr){8-10}
 & $c=39$ & $c=78$ & $c=130$ & $c=39$ & $c=78$ & $c=130$ & $c=39$ & $c=78$ & $c=130$ \\
\midrule
\textbf{DH} & 0.037 & 0.039 & 0.041 & \textbf{0.023} & 0.022 & 0.019 & 0.015 & 0.011 & 0.007\\
\textbf{t-iid} & 0.043 & 0.045 & 0.047 & 0.020 & 0.023 & 0.022 & 0.007 & 0.009 & 0.008\\
\textbf{t-MA} & 0.047 & \textbf{0.051} & \underline{0.052} & 0.021 & \textbf{0.025} & \underline{0.026} & 0.007 & \textbf{0.010} & 0.008\\
\textbf{t-iid (5)} & 0.062 & 0.063 & 0.059 & 0.029 & 0.032 & 0.035 & \underline{0.012} & 0.015 & 0.016\\
\textbf{t-MA (5)} & 0.065 & 0.067 & 0.066 & 0.028 & 0.034 & 0.035 & \textbf{0.011} & 0.014 & 0.017\\
\textbf{t-iid (21)} & \underline{0.047} & \underline{0.048} & \textbf{0.049} & 0.022 & \underline{0.025} & \textbf{0.026} & 0.008 & \underline{0.011} & \underline{0.009}\\
\textbf{t-MA (21)} & \textbf{0.050} & 0.055 & 0.056 & \underline{0.023} & 0.027 & 0.026 & 0.007 & 0.012 & \textbf{0.009}\\
\bottomrule
\end{tabular}

\vspace{0.1cm}

\begin{tabular}{lccccccccc}
\toprule
\multicolumn{10}{c}{\textbf{STUDENT'S T SYNTHETIC DATASET}} \\
\toprule
\multirow{2}{*}{\textbf{Algorithm}} & \multicolumn{3}{c}{$\theta=0.05$} & \multicolumn{3}{c}{$\theta=0.025$} & \multicolumn{3}{c}{$\theta=0.01$} \\
\cmidrule(lr){2-4} \cmidrule(lr){5-7} \cmidrule(lr){8-10}
 & $c=39$ & $c=78$ & $c=130$ & $c=39$ & $c=78$ & $c=130$ & $c=39$ & $c=78$ & $c=130$ \\
\midrule
\textbf{DH} & 0.071 & 0.095 & 0.116 & 0.019 & 0.035 & 0.043 & 0.001 & 0.001 & \textbf{0.010}\\
\textbf{t-iid} & \textbf{0.058} & \textbf{0.065} & \textbf{0.066} & \textbf{0.025} & \textbf{0.031} & \underline{0.035} & 0.008 & \underline{0.011} & \underline{0.011}\\
\textbf{t-MA} & 0.064 & 0.071 & \underline{0.067} & 0.027 & 0.034 & 0.038 & \underline{0.008} & 0.013 & 0.011\\
\textbf{t-iid (5)} & 0.080 & 0.088 & 0.082 & 0.039 & 0.046 & 0.043 & 0.015 & 0.017 & 0.016\\
\textbf{t-MA (5)} & 0.086 & 0.093 & 0.087 & 0.040 & 0.047 & 0.044 & 0.013 & 0.016 & 0.018\\
\textbf{t-iid (21)} & \underline{0.063} & \underline{0.069} & 0.067 & \underline{0.026} & \underline{0.032} & \textbf{0.032} & 0.008 & \textbf{0.010} & 0.011\\
\textbf{t-MA (21)} & 0.067 & 0.074 & 0.071 & 0.029 & 0.036 & 0.035 & \textbf{0.009} & 0.012 & 0.011\\
\bottomrule
\end{tabular}

\caption{Synthetic dataset - hits frequency.}
\label{tab:sim_q_intests}
}
\end{center}
\end{table}

\begin{table}
\begin{center}
\begin{adjustwidth}{-0.2cm}{}
{\small

\begin{tabular}{lcccccccccccc}
\toprule
\multicolumn{13}{c}{\textbf{GAUSSIAN SYNTHETIC DATASET}} \\
\toprule
\multirow{4}{*}{\textbf{Algorithm}} & \multicolumn{6}{c}{$\theta=0.05$} & \multicolumn{6}{c}{$\theta=0.025$} \\
\cmidrule(lr){2-7} \cmidrule(lr){8-13}
 & \multicolumn{2}{c}{$c=39$} & \multicolumn{2}{c}{$c=78$} & \multicolumn{2}{c}{$c=130$} & \multicolumn{2}{c}{$c=39$} & \multicolumn{2}{c}{$c=78$} & \multicolumn{2}{c}{$c=130$} \\
\cmidrule(lr){2-3} \cmidrule(lr){4-5} \cmidrule(lr){6-7} \cmidrule(lr){8-9} \cmidrule(lr){10-11} \cmidrule(lr){12-13}
 & AS1 & AS2 & AS1 & AS2 & AS1 & AS2 & AS1 & AS2 & AS1 & AS2 & AS1 & AS2 \\
\midrule
\textbf{DH} & \textbf{0.000} & 0.250 & 0.200 & 0.100 & 0.300 & 0.100 & 0.050 & \textbf{0.000} & \textbf{0.000} & \textbf{0.000} & 0.150 & 0.100\\
\textbf{t-iid} & 0.050 & \underline{0.050} & 0.050 & 0.050 & 0.300 & \textbf{0.000} & 0.150 & 0.050 & 0.050 & \textbf{0.000} & 0.200 & 0.050\\
\textbf{t-MA} & 0.100 & \underline{0.050} & 0.100 & \textbf{0.000} & 0.200 & \textbf{0.000} & 0.150 & \textbf{0.000} & 0.200 & \textbf{0.000} & 0.300 & 0.050\\
\textbf{t-iid (5)} & 0.050 & 0.100 & \textbf{0.000} & 0.100 & \underline{0.050} & 0.050 & \textbf{0.000} & \textbf{0.000} & \textbf{0.000} & 0.050 & 0.100 & \textbf{0.000}\\
\textbf{t-MA (5)} & \textbf{0.000} & \underline{0.050} & \textbf{0.000} & 0.100 & \textbf{0.000} & 0.100 & \textbf{0.000} & \textbf{0.000} & \textbf{0.000} & \textbf{0.000} & \textbf{0.000} & \textbf{0.000}\\
\textbf{t-iid (21)} & 0.050 & \textbf{0.000} & 0.100 & \textbf{0.000} & \underline{0.050} & \textbf{0.000} & 0.150 & 0.050 & 0.100 & \textbf{0.000} & \underline{0.050} & 0.050\\
\textbf{t-MA (21)} & 0.050 & \underline{0.050} & 0.050 & \textbf{0.000} & \underline{0.050} & \textbf{0.000} & 0.100 & \textbf{0.000} & 0.100 & \textbf{0.000} & \underline{0.050} & \textbf{0.000}\\
\bottomrule
\end{tabular}

\vspace{0.1cm}

\begin{tabular}{lcccccccccccc}
\toprule
\multicolumn{13}{c}{\textbf{STUDENT'S T SYNTHETIC DATASET}} \\
\toprule
\multirow{4}{*}{\textbf{Algorithm}} & \multicolumn{6}{c}{$\theta=0.05$} & \multicolumn{6}{c}{$\theta=0.025$} \\
\cmidrule(lr){2-7} \cmidrule(lr){8-13}
 & \multicolumn{2}{c}{$c=39$} & \multicolumn{2}{c}{$c=78$} & \multicolumn{2}{c}{$c=130$} & \multicolumn{2}{c}{$c=39$} & \multicolumn{2}{c}{$c=78$} & \multicolumn{2}{c}{$c=130$} \\
\cmidrule(lr){2-3} \cmidrule(lr){4-5} \cmidrule(lr){6-7} \cmidrule(lr){8-9} \cmidrule(lr){10-11} \cmidrule(lr){12-13}
 & AS1 & AS2 & AS1 & AS2 & AS1 & AS2 & AS1 & AS2 & AS1 & AS2 & AS1 & AS2 \\
\midrule
\textbf{DH} & 1.000 & 0.450 & 1.000 & 0.150 & 1.000 & 0.150 & 0.400 & 0.450 & 0.750 & 0.400 & 1.000 & 0.250\\
\textbf{t-iid} & \underline{0.150} & \textbf{0.050} & \textbf{0.050} & \textbf{0.100} & 0.050 & 0.050 & 0.150 & \textbf{0.000} & 0.250 & 0.050 & 0.250 & \textbf{0.000}\\
\textbf{t-MA} & \underline{0.150} & \textbf{0.050} & 0.100 & 0.150 & 0.050 & 0.050 & 0.200 & \textbf{0.000} & 0.150 & 0.050 & 0.400 & \textbf{0.000}\\
\textbf{t-iid (5)} & \textbf{0.050} & 0.100 & 0.150 & 0.350 & \textbf{0.000} & 0.150 & \textbf{0.050} & 0.050 & 0.150 & \textbf{0.000} & \textbf{0.100} & \textbf{0.000}\\
\textbf{t-MA (5)} & 0.250 & 0.100 & 0.200 & 0.350 & 0.050 & 0.150 & 0.300 & \textbf{0.000} & 0.300 & \textbf{0.000} & \textbf{0.100} & \textbf{0.000}\\
\textbf{t-iid (21)} & \underline{0.150} & \textbf{0.050} & \textbf{0.050} & \textbf{0.100} & \textbf{0.000} & \textbf{0.000} & \underline{0.100} & \textbf{0.000} & \textbf{0.100} & \textbf{0.000} & 0.150 & \textbf{0.000}\\
\textbf{t-MA (21)} & 0.200 & \textbf{0.050} & 0.150 & 0.150 & 0.050 & \textbf{0.000} & 0.200 & \textbf{0.000} & \textbf{0.100} & \textbf{0.000} & 0.250 & \textbf{0.000}\\
\bottomrule
\end{tabular}

\caption{Synthetic dataset - AS1 and AS2 in-sample tests.}
\label{tab:sim_e_intests}
}
\end{adjustwidth}
\end{center}
\end{table}

\clearpage
\section{Further Results on Forecasting Realized Risk Measures}
\label{app:for_rrm}
In this appendix, we show further results for the out-of-sample validation of the proposed methodology. Table \ref{tab:real_loss_q_ema} shows the Pinball loss when the predictor is an Exponential Moving Average (EMA); Table \ref{tab:real_loss_e_ema} shows the results for the ES. The EMA equation for estimated VaR reads:
\begin{equation}
    EMA_{t} = \alpha \cdot \hat{q}_{t-1} + (1-\alpha) \cdot EMA_{t-1}
\end{equation}
where the coefficient $\alpha$ is set to 0.9, as standard in the literature. The equation for the estimated ES is similar. The train set covers one year.
\begin{table}[h]
\begin{center}
{\small

\begin{tabular}{lccccccccc}
\toprule
\multicolumn{10}{c}{\textbf{BANK DATASET}} \\
\toprule
\multirow{2}{*}{\textbf{Algorithm}} & \multicolumn{3}{c}{$\theta=0.05$} & \multicolumn{3}{c}{$\theta=0.025$} & \multicolumn{3}{c}{$\theta=0.01$} \\
\cmidrule(lr){2-4} \cmidrule(lr){5-7} \cmidrule(lr){8-10}
 & $c=39$ & $c=78$ & $c=130$ & $c=39$ & $c=78$ & $c=130$ & $c=39$ & $c=78$ & $c=130$ \\
\midrule
\textbf{DH (Clock)} & 1.953 & 1.881 & 1.859 & 1.215 & 1.145 & 1.118 & 0.658 & 0.587 & 0.565\\
\textbf{DH (TPV)} & 1.909 & 1.865 & \textbf{1.842} & 1.186 & 1.127 & 1.111 & 0.656 & 0.592 & 0.57\\
\textbf{DH (Vol)} & 1.925 & 1.859 & \underline{1.844} & 1.182 & 1.126 & 1.109 & 0.642 & 0.589 & 0.564\\
\midrule
\textbf{t-iid (Clock)} & \underline{1.858} & \textbf{1.844} & 1.845 & \underline{1.111} & \textbf{1.099} & \textbf{1.094} & 0.56 & \underline{0.549} & \textbf{0.541}\\
\textbf{t-iid (TPV)} & 1.858 & 1.866 & 1.899 & 1.116 & 1.116 & 1.131 & 0.564 & 0.564 & 0.563\\
\textbf{t-iid (Vol)} & 1.877 & 1.89 & 1.926 & 1.125 & 1.13 & 1.151 & 0.567 & 0.565 & 0.577\\
\midrule
\textbf{t-MA (Clock)} & 1.865 & \underline{1.856} & 1.866 & 1.119 & 1.107 & 1.107 & 0.561 & \textbf{0.548} & 0.543\\
\textbf{t-MA (TPV)} & \textbf{1.853} & 1.883 & 1.932 & \textbf{1.107} & 1.135 & 1.156 & 0.559 & 0.576 & 0.577\\
\textbf{t-MA (Vol)} & 1.874 & 1.898 & 1.946 & 1.121 & 1.138 & 1.163 & 0.559 & 0.566 & 0.581\\
\midrule
\textbf{t-iid (5; Clock)} & 1.976 & 1.958 & 1.956 & 1.175 & 1.161 & 1.157 & 0.579 & 0.569 & 0.566\\
\textbf{t-iid (5; TPV)} & 1.959 & 1.962 & 1.988 & 1.173 & 1.17 & 1.179 & 0.589 & 0.588 & 0.579\\
\textbf{t-iid (5; Vol)} & 1.973 & 1.983 & 2.009 & 1.176 & 1.182 & 1.195 & 0.591 & 0.587 & 0.589\\
\midrule
\textbf{t-MA (5; Clock)} & 1.937 & 1.938 & 1.932 & 1.153 & 1.15 & 1.141 & 0.572 & 0.565 & 0.557\\
\textbf{t-MA (5; TPV)} & 1.932 & 1.955 & 1.993 & 1.154 & 1.173 & 1.186 & 0.577 & 0.584 & 0.589\\
\textbf{t-MA (5; Vol)} & 1.944 & 1.96 & 1.996 & 1.156 & 1.167 & 1.184 & 0.572 & 0.576 & 0.58\\
\midrule
\textbf{t-iid (21; Clock)} & 1.88 & 1.861 & 1.861 & 1.122 & \underline{1.107} & \underline{1.102} & 0.562 & 0.55 & \underline{0.543}\\
\textbf{t-iid (21; TPV)} & 1.867 & 1.879 & 1.916 & 1.12 & 1.123 & 1.137 & 0.562 & 0.566 & 0.563\\
\textbf{t-iid (21; Vol)} & 1.892 & 1.903 & 1.938 & 1.13 & 1.134 & 1.156 & 0.568 & 0.565 & 0.577\\
\midrule
\textbf{t-MA (21; Clock)} & 1.876 & 1.872 & 1.88 & 1.12 & 1.112 & 1.12 & 0.561 & 0.55 & 0.55\\
\textbf{t-MA (21; TPV)} & 1.865 & 1.897 & 1.94 & 1.114 & 1.137 & 1.16 & \underline{0.559} & 0.572 & 0.579\\
\textbf{t-MA (21; Vol)} & 1.881 & 1.907 & 1.954 & 1.122 & 1.139 & 1.17 & \textbf{0.558} & 0.563 & 0.577\\
\bottomrule
\end{tabular}

\vspace{0.1cm}

\begin{tabular}{lccccccccc}
\toprule
\multicolumn{10}{c}{\textbf{NASDAQ DATASET}} \\
\toprule
\multirow{2}{*}{\textbf{Algorithm}} & \multicolumn{3}{c}{$\theta=0.05$} & \multicolumn{3}{c}{$\theta=0.025$} & \multicolumn{3}{c}{$\theta=0.01$} \\
\cmidrule(lr){2-4} \cmidrule(lr){5-7} \cmidrule(lr){8-10}
 & $c=39$ & $c=78$ & $c=130$ & $c=39$ & $c=78$ & $c=130$ & $c=39$ & $c=78$ & $c=130$ \\
\midrule
\textbf{DH (Clock)} & 2.04 & 1.976 & 1.949 & 1.259 & 1.199 & 1.162 & 0.671 & 0.603 & 0.579\\
\textbf{DH (TPV)} & 1.988 & \textbf{1.937} & \textbf{1.921} & 1.224 & 1.167 & \textbf{1.146} & 0.665 & 0.597 & 0.574\\
\textbf{DH (Vol)} & 1.996 & \underline{1.94} & \underline{1.93} & 1.222 & 1.161 & \underline{1.148} & 0.636 & 0.593 & 0.575\\
\midrule
\textbf{t-iid (Clock)} & 1.952 & 1.953 & 1.943 & 1.167 & 1.162 & 1.151 & 0.584 & 0.576 & \underline{0.563}\\
\textbf{t-iid (TPV)} & \textbf{1.935} & 1.945 & 1.999 & \textbf{1.146} & \textbf{1.152} & 1.184 & \textbf{0.564} & \textbf{0.565} & 0.577\\
\textbf{t-iid (Vol)} & 1.953 & 1.985 & 2.043 & 1.162 & 1.183 & 1.217 & 0.574 & 0.586 & 0.598\\
\midrule
\textbf{t-MA (Clock)} & 1.951 & 1.949 & 1.953 & 1.164 & 1.157 & 1.153 & 0.577 & 0.571 & \textbf{0.563}\\
\textbf{t-MA (TPV)} & \underline{1.946} & 1.953 & 2.026 & \underline{1.152} & \underline{1.156} & 1.199 & \underline{0.566} & \underline{0.568} & 0.582\\
\textbf{t-MA (Vol)} & 1.957 & 1.997 & 2.078 & 1.158 & 1.188 & 1.24 & 0.57 & 0.583 & 0.609\\
\midrule
\textbf{t-iid (5; Clock)} & 2.08 & 2.074 & 2.062 & 1.242 & 1.237 & 1.222 & 0.617 & 0.609 & 0.599\\
\textbf{t-iid (5; TPV)} & 2.041 & 2.059 & 2.089 & 1.215 & 1.225 & 1.235 & 0.601 & 0.602 & 0.601\\
\textbf{t-iid (5; Vol)} & 2.064 & 2.084 & 2.116 & 1.231 & 1.243 & 1.253 & 0.61 & 0.616 & 0.614\\
\midrule
\textbf{t-MA (5; Clock)} & 2.051 & 2.039 & 2.047 & 1.22 & 1.208 & 1.209 & 0.606 & 0.591 & 0.588\\
\textbf{t-MA (5; TPV)} & 2.03 & 2.045 & 2.083 & 1.205 & 1.208 & 1.228 & 0.592 & 0.591 & 0.595\\
\textbf{t-MA (5; Vol)} & 2.048 & 2.061 & 2.129 & 1.21 & 1.223 & 1.264 & 0.592 & 0.601 & 0.621\\
\midrule
\textbf{t-iid (21; Clock)} & 1.985 & 1.982 & 1.967 & 1.187 & 1.181 & 1.169 & 0.592 & 0.587 & 0.574\\
\textbf{t-iid (21; TPV)} & 1.962 & 1.972 & 2.019 & 1.166 & 1.172 & 1.194 & 0.577 & 0.578 & 0.584\\
\textbf{t-iid (21; Vol)} & 1.982 & 2.007 & 2.052 & 1.181 & 1.198 & 1.223 & 0.586 & 0.594 & 0.602\\
\midrule
\textbf{t-MA (21; Clock)} & 1.976 & 1.973 & 1.977 & 1.18 & 1.171 & 1.168 & 0.586 & 0.578 & 0.571\\
\textbf{t-MA (21; TPV)} & 1.968 & 1.977 & 2.035 & 1.167 & 1.173 & 1.204 & 0.575 & 0.578 & 0.586\\
\textbf{t-MA (21; Vol)} & 1.987 & 2.01 & 2.085 & 1.177 & 1.196 & 1.243 & 0.579 & 0.589 & 0.611\\
\bottomrule
\end{tabular}

\caption{Real-world dataset - quantile loss function for forecasting. The predictor model is an Exponential Moving Average (EMA).}
\label{tab:real_loss_q_ema}
}
\end{center}
\end{table}
\begin{table}[h]
\begin{center}
\begin{adjustwidth}{-1.2cm}{}
{\small

\begin{tabular}{lcccccc||cccccc}
\toprule
 & \multicolumn{6}{c}{\textbf{BANK DATASET}} & \multicolumn{6}{c}{\textbf{NASDAQ DATASET}} \\
\cmidrule(lr){2-7} \cmidrule(lr){8-13}
\multirow{2}{*}{\textbf{Algorithm}} & \multicolumn{3}{c}{$\theta=0.05$} & \multicolumn{3}{c}{$\theta=0.025$} & \multicolumn{3}{c}{$\theta=0.05$} & \multicolumn{3}{c}{$\theta=0.025$} \\
\cmidrule(lr){2-4} \cmidrule(lr){5-7} \cmidrule(lr){8-10} \cmidrule(lr){11-13}
 & $c=39$ & $c=78$ & $c=130$ & $c=39$ & $c=78$ & $c=130$ & $c=39$ & $c=78$ & $c=130$ & $c=39$ & $c=78$ & $c=130$ \\
\midrule
\textbf{DH (Clock)} & 1.304 & 1.214 & 1.19 & 1.594 & 1.435 & 1.39 & 1.386 & 1.298 & 1.267 & 1.668 & 1.514 & 1.456\\
\textbf{DH (TPV)} & 1.261 & 1.198 & 1.183 & 1.548 & 1.422 & 1.394 & 1.335 & 1.27 & 1.26 & 1.62 & 1.492 & 1.457\\
\textbf{DH (Vol)} & 1.263 & 1.204 & 1.194 & 1.53 & 1.425 & 1.398 & 1.331 & 1.274 & 1.271 & 1.586 & 1.484 & 1.467\\
\midrule
\textbf{t-iid (Clock)} & 1.191 & \underline{1.176} & \textbf{1.168} & 1.392 & \underline{1.37} & \textbf{1.353} & 1.272 & 1.263 & \textbf{1.25} & 1.468 & 1.452 & \underline{1.431}\\
\textbf{t-iid (TPV)} & \underline{1.181} & 1.177 & 1.188 & 1.38 & 1.371 & 1.374 & \textbf{1.257} & \textbf{1.256} & 1.273 & \underline{1.446} & \textbf{1.44} & 1.448\\
\textbf{t-iid (Vol)} & 1.188 & 1.19 & 1.211 & 1.385 & 1.38 & 1.398 & 1.264 & 1.276 & 1.301 & 1.454 & 1.462 & 1.482\\
\midrule
\textbf{t-MA (Clock)} & 1.183 & \textbf{1.173} & \underline{1.174} & 1.375 & \textbf{1.36} & \underline{1.357} & 1.263 & \underline{1.257} & \underline{1.251} & 1.452 & \underline{1.441} & \textbf{1.428}\\
\textbf{t-MA (TPV)} & \textbf{1.173} & 1.185 & 1.2 & \textbf{1.363} & 1.379 & 1.384 & \underline{1.26} & 1.259 & 1.287 & 1.447 & 1.442 & 1.459\\
\textbf{t-MA (Vol)} & 1.181 & 1.193 & 1.219 & \underline{1.371} & 1.381 & 1.405 & 1.261 & 1.281 & 1.32 & \textbf{1.443} & 1.463 & 1.502\\
\midrule
\textbf{t-iid (5; Clock)} & 1.337 & 1.304 & 1.29 & 1.528 & 1.489 & 1.471 & 1.428 & 1.404 & 1.378 & 1.623 & 1.594 & 1.559\\
\textbf{t-iid (5; TPV)} & 1.306 & 1.297 & 1.276 & 1.505 & 1.492 & 1.458 & 1.391 & 1.386 & 1.354 & 1.593 & 1.579 & 1.531\\
\textbf{t-iid (5; Vol)} & 1.306 & 1.289 & 1.278 & 1.498 & 1.477 & 1.46 & 1.393 & 1.368 & 1.348 & 1.587 & 1.556 & 1.522\\
\midrule
\textbf{t-MA (5; Clock)} & 1.267 & 1.261 & 1.251 & 1.455 & 1.445 & 1.431 & 1.376 & 1.346 & 1.338 & 1.561 & 1.53 & 1.516\\
\textbf{t-MA (5; TPV)} & 1.267 & 1.266 & 1.255 & 1.462 & 1.461 & 1.441 & 1.35 & 1.35 & 1.333 & 1.543 & 1.534 & 1.505\\
\textbf{t-MA (5; Vol)} & 1.258 & 1.256 & 1.255 & 1.44 & 1.441 & 1.437 & 1.349 & 1.331 & 1.348 & 1.529 & 1.515 & 1.526\\
\midrule
\textbf{t-iid (21; Clock)} & 1.224 & 1.198 & 1.19 & 1.425 & 1.392 & 1.377 & 1.313 & 1.298 & 1.278 & 1.513 & 1.49 & 1.464\\
\textbf{t-iid (21; TPV)} & 1.206 & 1.201 & 1.204 & 1.407 & 1.399 & 1.388 & 1.294 & 1.288 & 1.29 & 1.491 & 1.482 & 1.467\\
\textbf{t-iid (21; Vol)} & 1.211 & 1.206 & 1.22 & 1.406 & 1.395 & 1.407 & 1.298 & 1.297 & 1.305 & 1.494 & 1.487 & 1.488\\
\midrule
\textbf{t-MA (21; Clock)} & 1.202 & 1.191 & 1.192 & 1.395 & 1.38 & 1.379 & 1.295 & 1.284 & 1.274 & 1.487 & 1.471 & 1.452\\
\textbf{t-MA (21; TPV)} & 1.194 & 1.203 & 1.21 & 1.386 & 1.399 & 1.398 & 1.286 & 1.284 & 1.294 & 1.479 & 1.474 & 1.468\\
\textbf{t-MA (21; Vol)} & 1.195 & 1.203 & 1.224 & 1.383 & 1.39 & 1.41 & 1.292 & 1.292 & 1.322 & 1.477 & 1.478 & 1.504\\
\bottomrule
\end{tabular}

}
\end{adjustwidth}
\caption{Real-world dataset - out-of-sample ES loss function. The Exponential Moving Average (EMA) is used to forecast the future value of VaR and ES, given the current realized value.}
\label{tab:real_loss_e_ema}
\end{center}
\end{table}\\

A further analysis is carried out using the Random Walk predictor; that is, the risk measure predicted at time $t+1$ is the realized risk measure at time $t$. Table \ref{tab:real_loss_q_rw} shows the VaR results, and Table \ref{tab:real_loss_e_rw} is for ES. Overall, the qualitative results are the same as in the main text, thus confirming the superiority of the proposed estimation approach with respect to the state-of-the-art of Ref. \cite{dimitriadis2022realized}.
\begin{table}[h]
\begin{center}
{\small

\begin{tabular}{lccccccccc}
\toprule
\multicolumn{10}{c}{\textbf{BANK DATASET}} \\
\toprule
\multirow{2}{*}{\textbf{Algorithm}} & \multicolumn{3}{c}{$\theta=0.05$} & \multicolumn{3}{c}{$\theta=0.025$} & \multicolumn{3}{c}{$\theta=0.01$} \\
\cmidrule(lr){2-4} \cmidrule(lr){5-7} \cmidrule(lr){8-10}
 & $c=39$ & $c=78$ & $c=130$ & $c=39$ & $c=78$ & $c=130$ & $c=39$ & $c=78$ & $c=130$ \\
\midrule
\textbf{DH (Clock)} & 1.983 & 1.895 & 1.869 & 1.242 & 1.161 & 1.128 & 0.688 & 0.603 & 0.576\\
\textbf{DH (TPV)} & 1.93 & 1.878 & \textbf{1.851} & 1.206 & 1.139 & 1.119 & 0.676 & 0.606 & 0.581\\
\textbf{DH (Vol)} & 1.95 & 1.874 & 1.854 & 1.207 & 1.14 & 1.12 & 0.668 & 0.605 & 0.574\\
\midrule
\textbf{t-iid (Clock)} & 1.869 & \textbf{1.852} & \underline{1.851} & 1.119 & \textbf{1.108} & \textbf{1.099} & 0.568 & 0.556 & \underline{0.545}\\
\textbf{t-iid (TPV)} & \underline{1.863} & 1.871 & 1.908 & 1.119 & 1.119 & 1.138 & 0.566 & 0.566 & 0.567\\
\textbf{t-iid (Vol)} & 1.887 & 1.899 & 1.936 & 1.136 & 1.139 & 1.159 & 0.576 & 0.573 & 0.583\\
\midrule
\textbf{t-MA (Clock)} & 1.868 & \underline{1.859} & 1.868 & 1.121 & \underline{1.11} & \underline{1.108} & 0.564 & \textbf{0.551} & \textbf{0.544}\\
\textbf{t-MA (TPV)} & \textbf{1.855} & 1.885 & 1.934 & \textbf{1.109} & 1.137 & 1.157 & \textbf{0.559} & 0.577 & 0.578\\
\textbf{t-MA (Vol)} & 1.877 & 1.901 & 1.949 & 1.125 & 1.141 & 1.165 & 0.564 & 0.569 & 0.583\\
\midrule
\textbf{t-iid (5; Clock)} & 2.003 & 1.981 & 1.978 & 1.196 & 1.18 & 1.174 & 0.596 & 0.582 & 0.579\\
\textbf{t-iid (5; TPV)} & 1.974 & 1.98 & 2.01 & 1.185 & 1.183 & 1.196 & 0.597 & 0.596 & 0.591\\
\textbf{t-iid (5; Vol)} & 1.997 & 2.006 & 2.035 & 1.195 & 1.2 & 1.214 & 0.606 & 0.6 & 0.602\\
\midrule
\textbf{t-MA (5; Clock)} & 1.948 & 1.951 & 1.941 & 1.162 & 1.158 & 1.147 & 0.579 & 0.571 & 0.562\\
\textbf{t-MA (5; TPV)} & 1.941 & 1.963 & 2.001 & 1.159 & 1.179 & 1.192 & 0.581 & 0.588 & 0.594\\
\textbf{t-MA (5; Vol)} & 1.956 & 1.972 & 2.006 & 1.165 & 1.175 & 1.191 & 0.579 & 0.581 & 0.584\\
\midrule
\textbf{t-iid (21; Clock)} & 1.893 & 1.871 & 1.87 & 1.132 & 1.116 & 1.109 & 0.572 & 0.557 & 0.549\\
\textbf{t-iid (21; TPV)} & 1.873 & 1.886 & 1.927 & 1.124 & 1.127 & 1.145 & 0.565 & 0.569 & 0.567\\
\textbf{t-iid (21; Vol)} & 1.904 & 1.915 & 1.95 & 1.141 & 1.144 & 1.165 & 0.577 & 0.573 & 0.583\\
\midrule
\textbf{t-MA (21; Clock)} & 1.88 & 1.877 & 1.883 & 1.124 & 1.116 & 1.122 & 0.565 & \underline{0.553} & 0.552\\
\textbf{t-MA (21; TPV)} & 1.867 & 1.899 & 1.943 & \underline{1.116} & 1.139 & 1.162 & \underline{0.559} & 0.572 & 0.58\\
\textbf{t-MA (21; Vol)} & 1.886 & 1.912 & 1.958 & 1.127 & 1.143 & 1.172 & 0.562 & 0.566 & 0.58\\
\bottomrule
\end{tabular}

\vspace{0.1cm}

\begin{tabular}{lccccccccc}
\toprule
\multicolumn{10}{c}{\textbf{NASDAQ DATASET}} \\
\toprule
\multirow{2}{*}{\textbf{Algorithm}} & \multicolumn{3}{c}{$\theta=0.05$} & \multicolumn{3}{c}{$\theta=0.025$} & \multicolumn{3}{c}{$\theta=0.01$} \\
\cmidrule(lr){2-4} \cmidrule(lr){5-7} \cmidrule(lr){8-10}
 & $c=39$ & $c=78$ & $c=130$ & $c=39$ & $c=78$ & $c=130$ & $c=39$ & $c=78$ & $c=130$ \\
\midrule
\textbf{DH (Clock)} & 2.074 & 1.998 & 1.966 & 1.295 & 1.223 & 1.177 & 0.703 & 0.621 & 0.592\\
\textbf{DH (TPV)} & 2.016 & \underline{1.955} & \textbf{1.935} & 1.249 & 1.184 & \underline{1.158} & 0.689 & 0.612 & 0.585\\
\textbf{DH (Vol)} & 2.027 & 1.959 & \underline{1.944} & 1.25 & 1.178 & 1.158 & 0.662 & 0.608 & 0.585\\
\midrule
\textbf{t-iid (Clock)} & 1.969 & 1.967 & 1.954 & 1.181 & 1.176 & 1.162 & 0.595 & 0.586 & \underline{0.569}\\
\textbf{t-iid (TPV)} & \textbf{1.943} & \textbf{1.954} & 2.011 & \textbf{1.152} & \underline{1.159} & 1.193 & \textbf{0.568} & \underline{0.571} & 0.583\\
\textbf{t-iid (Vol)} & 1.963 & 1.997 & 2.055 & 1.171 & 1.193 & 1.226 & 0.582 & 0.593 & 0.604\\
\midrule
\textbf{t-MA (Clock)} & 1.958 & 1.956 & 1.957 & 1.17 & 1.162 & \textbf{1.156} & 0.582 & 0.575 & \textbf{0.564}\\
\textbf{t-MA (TPV)} & \underline{1.95} & 1.957 & 2.03 & \underline{1.156} & \textbf{1.159} & 1.202 & \underline{0.569} & \textbf{0.57} & 0.585\\
\textbf{t-MA (Vol)} & 1.963 & 2.003 & 2.082 & 1.163 & 1.193 & 1.244 & 0.574 & 0.587 & 0.612\\
\midrule
\textbf{t-iid (5; Clock)} & 2.111 & 2.103 & 2.087 & 1.266 & 1.257 & 1.241 & 0.635 & 0.625 & 0.612\\
\textbf{t-iid (5; TPV)} & 2.057 & 2.078 & 2.111 & 1.227 & 1.239 & 1.251 & 0.608 & 0.612 & 0.613\\
\textbf{t-iid (5; Vol)} & 2.084 & 2.106 & 2.139 & 1.246 & 1.26 & 1.268 & 0.622 & 0.628 & 0.624\\
\midrule
\textbf{t-MA (5; Clock)} & 2.066 & 2.053 & 2.058 & 1.231 & 1.216 & 1.216 & 0.615 & 0.596 & 0.592\\
\textbf{t-MA (5; TPV)} & 2.039 & 2.055 & 2.094 & 1.211 & 1.215 & 1.236 & 0.596 & 0.596 & 0.599\\
\textbf{t-MA (5; Vol)} & 2.059 & 2.072 & 2.139 & 1.218 & 1.23 & 1.269 & 0.598 & 0.607 & 0.625\\
\midrule
\textbf{t-iid (21; Clock)} & 2.002 & 1.998 & 1.98 & 1.201 & 1.194 & 1.178 & 0.604 & 0.597 & 0.582\\
\textbf{t-iid (21; TPV)} & 1.97 & 1.981 & 2.031 & 1.171 & 1.18 & 1.204 & 0.581 & 0.584 & 0.591\\
\textbf{t-iid (21; Vol)} & 1.994 & 2.02 & 2.064 & 1.191 & 1.209 & 1.233 & 0.594 & 0.602 & 0.609\\
\midrule
\textbf{t-MA (21; Clock)} & 1.984 & 1.979 & 1.982 & 1.187 & 1.177 & 1.171 & 0.591 & 0.582 & 0.573\\
\textbf{t-MA (21; TPV)} & 1.973 & 1.98 & 2.04 & 1.17 & 1.176 & 1.207 & 0.578 & 0.58 & 0.588\\
\textbf{t-MA (21; Vol)} & 1.994 & 2.015 & 2.091 & 1.182 & 1.201 & 1.247 & 0.583 & 0.593 & 0.614\\
\bottomrule
\end{tabular}

\caption{Real-world dataset - quantile loss function for forecasting. The predictor model is a Random Walk (RW).}
\label{tab:real_loss_q_rw}
}
\end{center}
\end{table}\\
\begin{table}[h]
\begin{center}
\begin{adjustwidth}{-1.3cm}{}
{\small

\begin{tabular}{lcccccc||cccccc}
\toprule
 & \multicolumn{6}{c}{\textbf{BANK DATASET}} & \multicolumn{6}{c}{\textbf{NASDAQ DATASET}} \\
\cmidrule(lr){2-7} \cmidrule(lr){8-13}
\multirow{2}{*}{\textbf{Algorithm}} & \multicolumn{3}{c}{$\theta=0.05$} & \multicolumn{3}{c}{$\theta=0.025$} & \multicolumn{3}{c}{$\theta=0.05$} & \multicolumn{3}{c}{$\theta=0.025$} \\
\cmidrule(lr){2-4} \cmidrule(lr){5-7} \cmidrule(lr){8-10} \cmidrule(lr){11-13}
 & $c=39$ & $c=78$ & $c=130$ & $c=39$ & $c=78$ & $c=130$ & $c=39$ & $c=78$ & $c=130$ & $c=39$ & $c=78$ & $c=130$ \\
\midrule
\textbf{DH (Clock)} & 1.353 & 1.235 & 1.204 & 1.68 & 1.472 & 1.413 & 1.434 & 1.322 & 1.282 & 1.757 & 1.556 & 1.479\\
\textbf{DH (TPV)} & 1.292 & 1.215 & 1.195 & 1.605 & 1.451 & 1.414 & 1.37 & 1.288 & 1.272 & 1.683 & 1.523 & 1.476\\
\textbf{DH (Vol)} & 1.301 & 1.225 & 1.207 & 1.601 & 1.461 & 1.419 & 1.367 & 1.292 & 1.281 & 1.648 & 1.513 & 1.482\\
\midrule
\textbf{t-iid (Clock)} & 1.207 & 1.26 & \underline{1.186} & 1.415 & 1.502 & \underline{1.381} & 1.289 & 1.277 & \underline{1.259} & 1.495 & 1.475 & \underline{1.444}\\
\textbf{t-iid (TPV)} & 1.187 & \underline{1.181} & 1.195 & 1.388 & \underline{1.377} & 1.383 & \underline{1.264} & \underline{1.263} & 1.28 & 1.456 & 1.451 & 1.458\\
\textbf{t-iid (Vol)} & 1.201 & 1.2 & 1.22 & 1.405 & 1.396 & 1.41 & 1.273 & 1.285 & 1.307 & 1.469 & 1.475 & 1.491\\
\midrule
\textbf{t-MA (Clock)} & 1.187 & \textbf{1.177} & \textbf{1.177} & 1.383 & \textbf{1.366} & \textbf{1.361} & 1.271 & 1.263 & \textbf{1.254} & 1.464 & \underline{1.449} & \textbf{1.432}\\
\textbf{t-MA (TPV)} & \textbf{1.176} & 1.188 & 1.201 & \textbf{1.367} & 1.383 & 1.387 & \textbf{1.264} & \textbf{1.262} & 1.288 & \underline{1.453} & \textbf{1.445} & 1.462\\
\textbf{t-MA (Vol)} & \underline{1.186} & 1.196 & 1.222 & \underline{1.38} & 1.387 & 1.409 & 1.266 & 1.285 & 1.321 & \textbf{1.45} & 1.469 & 1.504\\
\midrule
\textbf{t-iid (5; Clock)} & 1.391 & 1.347 & 1.329 & 1.59 & 1.537 & 1.514 & 1.484 & 1.45 & 1.415 & 1.688 & 1.649 & 1.6\\
\textbf{t-iid (5; TPV)} & 1.337 & 1.35 & 1.309 & 1.54 & 1.526 & 1.496 & 1.417 & 1.415 & 1.383 & 1.622 & 1.611 & 1.562\\
\textbf{t-iid (5; Vol)} & 1.353 & 1.33 & 1.311 & 1.555 & 1.524 & 1.497 & 1.428 & 1.399 & 1.374 & 1.631 & 1.594 & 1.547\\
\midrule
\textbf{t-MA (5; Clock)} & 1.291 & 1.28 & 1.266 & 1.485 & 1.467 & 1.447 & 1.402 & 1.363 & 1.352 & 1.591 & 1.546 & 1.585\\
\textbf{t-MA (5; TPV)} & 1.282 & 1.281 & 1.266 & 1.477 & 1.482 & 1.455 & 1.365 & 1.363 & 1.343 & 1.56 & 1.549 & 1.517\\
\textbf{t-MA (5; Vol)} & 1.278 & 1.275 & 1.267 & 1.465 & 1.46 & 1.45 & 1.365 & 1.343 & 1.356 & 1.548 & 1.529 & 1.534\\
\midrule
\textbf{t-iid (21; Clock)} & 1.242 & 1.213 & 1.202 & 1.452 & 1.414 & 1.393 & 1.336 & 1.317 & 1.291 & 1.547 & 1.516 & 1.481\\
\textbf{t-iid (21; TPV)} & 1.214 & 1.208 & 1.213 & 1.419 & 1.408 & 1.400 & 1.303 & 1.298 & 1.299 & 1.504 & 1.496 & 1.480\\
\textbf{t-iid (21; Vol)} & 1.228 & 1.219 & 1.231 & 1.431 & 1.414 & 1.423 & 1.313 & 1.309 & 1.313 & 1.514 & 1.505 & 1.499\\
\midrule
\textbf{t-MA (21; Clock)} & 1.209 & 1.199 & 1.195 & 1.405 & 1.391 & 1.385 & 1.306 & 1.291 & 1.278 & 1.503 & 1.481 & 1.457\\
\textbf{t-MA (21; TPV)} & 1.197 & 1.206 & 1.213 & 1.392 & 1.404 & 1.401 & 1.291 & 1.288 & 1.297 & 1.486 & 1.48 & 1.471\\
\textbf{t-MA (21; Vol)} & 1.202 & 1.209 & 1.227 & 1.393 & 1.399 & 1.415 & 1.298 & 1.298 & 1.325 & 1.486 & 1.487 & 1.507\\
\midrule
\bottomrule
\end{tabular}

}
\end{adjustwidth}
\caption{Real-world dataset - out-of-sample ES loss function. The future values of VaR and ES are forecasted via a Random Walk (RW) model.}
\label{tab:real_loss_e_rw}
\end{center}
\end{table}
\end{document}